\documentclass[aps,showpacs,preprintnumbers,nofootinbibt,superscriptaddress, twocolumn, ]{revtex4-1}
\usepackage{eurosym}
\usepackage{graphicx}
\usepackage{amssymb}
\usepackage{amsmath}

\newcommand{\bea}{\begin{eqnarray}}
\newcommand{\eea}{\end{eqnarray}}
\newcommand{\beaa}{\begin{align}}
\newcommand{\eeaa}{\end{align}}
\newcommand{\be}{\begin{equation}}
\newcommand{\ee}{\end{equation}}

\begin{document}

\title{Palatini formulation of the conformally invariant $f\left(R,L_m\right)$ gravity theory }
\author{Tiberiu Harko}
\email{tiberiu.harko@aira.astro.ro}
\affiliation{Department of Theoretical Physics, National Institute of Physics
and Nuclear Engineering (IFIN-HH), Bucharest, 077125 Romania,} %
\affiliation{Department of Physics, Babes-Bolyai University, 1 Kogalniceanu Street,
400084 Cluj-Napoca, Romania,}
\affiliation{Astronomical Observatory, 19 Ciresilor Street, 400487
Cluj-Napoca, Romania,}
\author{Shahab Shahidi}
\email{s.shahidi@du.ac.ir}
\affiliation{School of Physics, Damghan University, Damghan, 36716-41167, Iran}
\begin{abstract}
We investigate the field equations of the conformally invariant models of gravity with curvature-matter coupling, constructed in Weyl geometry, by using the Palatini formalism. We consider the case in which the Lagrangian is given by the sum of the square of the Weyl scalar, of the strength of the field associated to the Weyl vector, and a conformally invariant geometry-matter coupling term, constructed from the matter Lagrangian and the Weyl scalar. After substituting the Weyl scalar in terms of its Riemannian counterpart, the quadratic action is defined in Riemann geometry, and involves a nonminimal coupling between the Ricci scalar and the matter Lagrangian. For the sake of generality, a more general Lagrangian, in which the Weyl vector is nonminmally coupled with an arbitrary function of the Ricci scalar, is also considered. By varying the action independently with respect to the metric and the connection, the independent connection can be expressed as the Levi-Civita connection of an auxiliary, Ricci scalar and Weyl vector dependent metric, which is related to the physical metric by means of a conformal transformation. The field equations are obtained in both the metric and the Palatini formulations. The cosmological implications of the Palatini field equations are investigated for three distinct models corresponding to different forms of the coupling functions. A comparison with the standard $\Lambda$CDM model is also performed, and we find that the Palatini type cosmological models can give an acceptable description of the observations.

\end{abstract}

\date{\today}
\maketitle

\tableofcontents

\section{Introduction}

The remarkable success of Einstein's theory of general relativity \cite{Eina}%
, and of its variational formulation \cite{Hil} gave a huge impetus not only
to gravitational physics, but also to mathematics. From a mathematical point
of view, general relativity is based on Riemannian geometry, with the metric
of the space-time describing all the properties of the gravitational
interaction. Almost immediately after the birth of general relativity, in an
attempt to unify the electromagnetic and gravitational interactions in a
fully geometric background, H. Weyl proposed a generalization of the Riemann
geometry by introducing a supplementary geometric quantity, called nonmetricity \cite%
{Weyl1,Weyl2,Weyl3,Weyl4,Weyl5}. The basic idea in Weyl's geometric approach
was to permit the length and orientation of arbitrary vectors to change
under parallel transport. In Riemannian geometry, only a change in the
orientation of the vectors is allowed. The geometry obtained in this way is called Weyl
geometry, and it represents a consistent and systematic generalization of
Riemannian geometry. In trying to provide a physical interpretation for his
geometry, Weyl identified the vector part of the connection with the potential four-vector
of the electromagnetic field. However, this identification is
problematic. Immediately after the publication of Weyl's theory, Einstein
\cite{Einb} pointed out that such an identification would imply that
identical atoms moving on closed trajectories in electromagnetic fields
would have different physical properties (and sizes), implying that the atoms would also have
different electromagnetic spectra. The change in frequency due to the size
change is clearly inconsistent with the well known observational properties of the
spectral lines. Einstein's criticism led to the abandonment of the
\textit{physical} Weyl theory in its initial formulation. However, later one,
Weyl proved that a satisfactory theory of the electromagnetic interaction is
obtained by replacing the scale factor with a complex phase. This remarkable
result is at the origin of U (1) gauge theory \cite{O}.

Even that Weyl's geometric theory failed in its attempt of unifying gravity
and electromagnetism, as a \textit{geometric} theory it has many attractive
features. It appears in many physical contexts, like, for example,
foundations of quantum mechanics, elementary particle physics, scalar tensor
theories of gravity, fundamental approaches to gravity, cosmology etc. Hence, Weyl
geometry has an important research potential, yet to be fully explored, which may
provide some deeper insights into the fundamental problems of gravitation, cosmology,
and elementary particle physics \cite{W1,W2,W3,W4,W5,W6}. Gravitational theories satisfying the requirement of conformal invariance can be obtained by using actions built up with the help of the Weyl tensor $C_{\alpha \beta \gamma \delta}$. The corresponding action is given by $S=-(1/4)\int{C_{\alpha \beta \gamma \delta}C^{\alpha \beta \gamma \delta}\sqrt{-g}d^4x}$ \cite{M1,M2,M3,M4, M5, M6}. Gravitational field theories described by actions of this type are called conformally invariant, or Weyl type gravity theories, and their properties and implications have been extensively investigated in the literature \cite{M1,M2,M3,M4, M5, M6}.

A particularly interesting theoretical issue is the possible relation
between the Standard Model of elementary particle physics, and of its
extensions, and Weyl geometry. The link between elementary particle physics
and geometry is provided by the concept of scale invariance, since this
symmetry may also appear at the quantum level \cite{Gh0, Gh01}. All physical
scales, including the new physical scales beyond the Standard Model, can be
generated spontaneously by vacuum expectation values of the fields.
Moreover, scale symmetry could maintain the classical hierarchy of scales
\cite{Sh1,Sh2}. The new vector part of the connection in Weyl geometry is
called, in the modern interpretation, the dilatational gauge vector, or the
Weyl vector.

A systematic study of the relation between the Standard Model (SM) of
elementary particle and Weyl conformal geometry, was initiated, and
performed, in \cite{Gh1,Gh2,Gh3,Gh4,Gh5,Gh6,Gh7}, by considering a minimal
embedding, with no new fields beyond the SM spectrum and Weyl geometry. The
action has the Weyl gauge symmetry D(1), originating from the background
geometry. The model is based on Weyl quadratic gravity, which experiences a
spontaneous breaking of D(1) symmetry by a geometric Stueckelberg mechanism,
with the Weyl gauge field acquiring mass from the spin-zero mode of the $R^2$
term in the action.

To describe the properties of the gravitational field, as well as physics
beyond SM, in \cite{Gh7} the following action was proposed
\begin{equation}  \label{S0}
S_{0}=\int \Big[\frac{1}{4!}\frac{1}{\xi ^{2}}\tilde{R}^{2}-\frac{1}{4}\,%
\tilde{F}_{\mu \nu }^{2}-\frac{1}{\eta ^{2}}\tilde{C}_{\mu \nu \rho \sigma
}^{2}\Big]\sqrt{-g}d^{4}x,
\end{equation}%
where $\tilde{R}$ is the Weyl scalar, defined in the Weyl geometry, $\tilde{F%
}_{\mu \nu }$ is the strength of the Weyl vector $\omega _{\mu}$, while $%
\tilde{C}_{\mu \nu \rho \sigma}$ denotes the conformally invariant Weyl
tensor.

In obtaining the gravitational field equations from a given action several
approaches can be used. The most common is the metric formalism, in which
the action is varied with respect to the metric tensor $g_{\mu \nu}$. One
can also use the Palatini formalism, introduced by Einstein \cite{Einc,Eind,
Eine}, where the metric and the connection $\Gamma$ are considered as
independent variables. To derive the field equations the Lagrangian is
varied with respect to both $g$ and $\Gamma$. Finally, in the metric-affine
formalism, which generalizes the Palatini approach, the matter part of the
action also depends on the connection, and it is varied with respect to it
\cite{Olmo1,Olmo2,Pal1,Pal2}.

In the case of the Einstein-Hilbert action, the Palatini
variation leads to the Einstein gravitational field equations, giving the
same result as when varying the metric only. However, this is not generally
true for other gravitational type actions. For example, when used for an $%
f(R)$ type gravitational Lagrangian, originally introduced in \cite{Bu70,Bu70a,Bu70b,Bu70c}, its metric version in the Palatini formalism leads to second
order differential equations instead of the fourth order ones obtained from
the metric variation \cite{P1,P2,P3,P4,P5,P6,P7,P8,P9,P10,P11,P12,P13}.
Moreover, in vacuum, the Palatini formalism $f(R)$ field equations reduce to
the field equations of standard General Relativity in the presence of a
cosmological constant. This result guarantees that the Palatini type $f(R)$
theory automatically passes the Solar System tests. Secondly, basic aspects
of general relativity, like the presence of black holes and of gravitational
waves, are preserved in the Palatini approach. However, up to present moment no real criterion has been found
indicating which variational formalism is better to apply. On the other
hand, the Palatini variational procedure seems to be more attractive, since, when applied to the Hilbert-Einstein action, one obtains
standard General Relativity without the need of specifying in advance the relation between metric
and connection.

Quadratic gravity with Lagrangian $R^2 + R_{[\mu \nu]}^2$ was investigated
in the Palatini formalism, by considering both the connection and the metric
independent, in \cite{Gh5}. The action has a gauged scale symmetry (or Weyl
gauge symmetry) with respect to the Weyl gauge field $\omega _{\mu }=\tilde{%
\Gamma}_{\mu }-\Gamma _{\mu }$, where $\tilde{\Gamma}_{\mu}$ and $\Gamma
_{\mu}$ are the traces of the Weyl and Levi-Civita connections,
respectively. In this approach the underlying geometry is non-metric due to
the presence of the $R_{[\mu \nu]}^2$ term acting as a gauge kinetic term
for $\omega _{\mu}$. This theory has a spontaneous breaking of gauged scale
symmetry and mass generation in the absence of matter. Interestingly, the
necessary scalar field $\varphi$ is not added ad-hoc, but it appears
naturally from the $R^2$ term. By absorbing the derivative term of the
Stueckelberg field the gauge field becomes massive. In the broken phase the
Einstein-Proca action of $\omega _{\mu}$ of mass proportional to the Planck
scale $M \sim <\varphi >$, and a positive cosmological constant are
obtained. Below the Planck scale $\omega _{\mu}$ decouples, the connection
becomes Levi-Civita, and the metricity condition and Einstein's general
relativity are recovered. These results remain valid in the presence of a
non-minimally coupled Higgs-like scalar field.

 A comparative study of inflation in two theories of quadratic
gravity with \textit{gauged} scale symmetry was performed in \cite{Gh5}. More exactly,
the original Weyl quadratic gravity, and a theory defined by a similar
action, but in the Palatini approach, obtained by replacing the Weyl
connection by its Palatini counterpart. These two theories have different
vectorial non-metricities, induced by the gauge field $\omega _{\mu}$. In
the absence of matter these theories have a spontaneous breaking of gauged
scale symmetry, where the necessary scalar field is of geometric origin, and
part of the quadratic action. The Einstein-Proca action, the Planck scale
and the metricity appear in the broken phase after the Weyl vector field
acquires mass through the Stueckelberg mechanism, and then decouples. In the
presence of nonminimally coupled matter, the scalar potential is similar in
both theories. For small field values the potential is Higgs-like, while for
large field values inflation is possible. Both theories have a small
tensor-to-scalar ratio $r\sim 10^{-3}$, slightly larger in the Palatini
case. For a small enough coupling parameter $\xi _1\leq 10^{-3}$, Weyl's
theory gives a dependence $r\left(n_s\right)$ similar to that in Starobinsky
inflation \cite{Sta1, Sta2, Sta3}.

 It was pointed out in \cite{Cai} that the model of the cosmic inflation with an asymptotically flat potential could be obtained from the Palatini quadratic gravity, formulated in Weyl geometry,  by adding the matter field in such a way that the local gauged conformal symmetry is broken in both kinetic and potential terms.

A possible way to explain the recent cosmological data indicating the presence
of an accelerated expansion of the Universe, and of the dark matter (for a review of the observational evidence for acceleration  see \cite{Wein}), is to postulate
that at galactic and extragalactic scales  Einstein's general theory of relativity must be replaced,
by a more general action that describes the gravitational phenomenology beyond the Solar System \cite{Od1,Od2}. Hence, observational evidence seems to strongly point out towards the necessity of going outside the strict limits of general relativity, and for looking to new gravitational theories that may solve (or account) for the dark energy and dark matter problems. Therefore, the Einstein gravitational field equations that provides a very good explanation of the gravitational processes in the Solar System, must be replaced by a new set of field equations.

A possible generalization of the standard Einstein theory
is represented by gravitational theories implying a curvature-matter coupling \cite{Tomi, Ha1,Ha2,Ha3,Ha4,Ha5,Ha6}. In this type of models the Hilbert-Einstein action of standard general relativity  $S=\int{\left(R/2\kappa ^2+L_m\right)\sqrt{-g}d^4x}$, where $R$ is the Ricci scalar, and $L_m$ is the matter Lagrangian, is replaced by more general actions of the form $S=\int{f\left(R,L_m\right)\sqrt{-g}d^4x}$ \cite{Ha2}, or $S=\int{f\left(R,T\right)\sqrt{-g}d^4x}$ \cite{Ha3}, where $T$ is the trace of the matter energy-momentum tensor. Similar couplings between matter and geometry are also possible in the presence of non-metricity and torsion \cite{Ha4,Ha5,Ha6}. For in depth
reviews and discussions of theories with curvature-matter coupling see \cite{Ha7,Ha8,Ha9, Ha10, Ha11}.
The curvature-matter coupling approach leads to gravitational theories having a much richer physical and mathematical structure as compared to standard general
relativity. They also provide interesting explanations for the accelerating expansion of the Universe, and possible solutions for the
dark energy and dark matter problems, respectively. But these types of theories also face a number of very difficult mathematical and physical problems. For the Palatini formulation of the $f\left(R,L_m\right)$ theory see \cite{Tomi1}.

For the $f(R, T )$ gravity theory \cite{Ha3}, in which a nonminimal coupling
between the Ricci scalar and the trace of the energy-momentum tensor is introduced, its Palatini formulation was  investigated in detail in \cite{Wu}, by considering the metric and the affine connection as independent field variables. For this type of theories, the independent connection can be expressed as the Levi-Civita connection of an auxiliary metric, depending on the trace of the energy-momentum tensor, and related to the physical metric by a conformal transformation. The field equations also lead to the non-conservation of the energy-momentum tensor. The thermodynamic interpretation of the Palatini formulation of the theory was also discussed. The cosmological implications of the theory have also been explored for several functional forms of the function $f$, and it was shown that the models can give an acceptable description of the observational cosmological data.

An investigation of the coupling between matter and geometry in conformal quadratic Weyl gravity was performed in \cite{HaSh}. To construct the physical model, and the gravitational action, a coupling term of the form $L_m\tilde{R}^2$ was assumed in the Lagrangian, where $\tilde{R}$ is the Weyl scalar. It is important to note that this  coupling explicitly satisfies the conformal invariance of the theory, under the assumption that the matter Lagrangian is conformally invariant. By expressing $\tilde{R}^ 2$ with the help of an auxiliary scalar field and of the Riemannian Ricci scalar, the gravitational action can be linearized, leading in the Riemann space to a conformally invariant $f \left(R,L_m\right)$ type theory, with the matter Lagrangian non-minimally coupled to the Ricci scalar.

The gravitational field equations of the theory can be  obtained in the metric formalism, together with the energy-momentum balance equations. Similarly to other theories with geometry-matter coupling the divergence of the matter energy-momentum tensor does not vanish, and an extra force, depending on the Weyl vector, and the matter Lagrangian does appear in the geodesic equations of motion. The theory can be interpreted thermodynamically as describing irreversible matter creation from the gravitational field. The generalized Poisson equation and the Newtonian limit of the equations of motion were also considered in detail. Constraints on the magnitude of the Weyl vector can be obtained from the study of the  perihelion precession of a planet in the presence of an extra force,  and an explicit estimation of the numerical value of a constant Weyl vector in the Solar System is obtained from the observational data of Mercury. The cosmological implications of the theory have been considered for the case of a flat, homogeneous and isotropic Friedmann-Lemaitre-Robertson-Walker geometry. It turns out that the conformally invariant $f\left(R,L_m\right)$ model gives a good description of the cosmological observational data for the Hubble function up to a redshift of the order of $z \approx 3$.

It is the goal of the present paper to consider the Palatini formulation of the conformally invariant $f\left(R,L_m\right)$ gravity theory proposed in \cite{HaSh}, which intrinsically contains a curvature-matter coupling. The gravitational action is constructed directly in the framework of Weyl geometry, with the Lagrangian density given by the sum of the square of the Weyl scalar $\tilde{R}$ (the analogue of the Ricci scalar in Weyl geometry), of the strength $F_{\mu \nu}$ of the geometric field associated to the Weyl vector $\omega _{\mu}$, and a conformally invariant geometry-matter coupling term $L_m\tilde{R}^2$, constructed from the matter Lagrangian and the Weyl scalar. After substituting the Weyl scalar $\tilde{R}$ in terms of its Riemannian counterpart $R$, one obtains a quadratic action defined in Riemann geometry, which contains a nonminimal coupling between the Ricci scalar, and the matter Lagrangian. {\it For the sake of generality}, we consider in our analysis {\it a more general Lagrangian, in which  the Weyl vector is nonminmally coupled to an arbitrary function of the Ricci scalar}. By varying the action independently with respect to the metric and the connection, it turns out that the {\it independent connection can be expressed as the Levi-Civita connection of an auxiliary, Ricci scalar and Weyl vector dependent metric}. This metric is related to the physical metric by means of a conformal transformation. We obtain the field equations in both the metric and the Palatini formulations. The cosmological implications of the Palatini field equations are investigated for three distinct theoretical models, corresponding to different (simple) forms of the coupling functions. The conformally invariant quadratic Weyl  model is investigated in detail. A comparison with the standard $\Lambda$CDM model is also performed. Our main findings indicates that the Palatini type cosmological models can give an acceptable description of the cosmological observations, at least up to a redshift of the order of $z\approx 2$.

The present paper is organized as follows. In Section~\ref{sect1}, after a brief review of the fundamentals of Weyl geometry, the gravitational action in the presence of conformally invariant, and an arbitrary geometry-matter coupling are written down, and the field equations are obtained by varying the action with respect to both metric and connection. The Palatini formulation of the conformally invariant $f\left(R,L_m\right)$ gravity theory is presented in detail in Section~\ref{sect2}. The cosmological implications of the Palatini formulation of the $f\left(R,L_m\right)$ gravity theory are presented in Section~\ref{sect3}. We discuss and conclude our results in Section~\ref{sect4}.

\section{Gravitational field equations in quadratic Weyl geometric gravity}\label{sect1}

In the present Section, after briefly reviewing the fundamentals of Weyl
geometry, we introduce the simplest conformally invariant gravitational
action in Weyl geometry in the presence of matter, constructed from the square of the Weyl scalar $\tilde{R}$, and
of the strengths of the Weyl vector $F_{\mu \nu}$. Moreover, we add to the gravitational action a geometry-matter coupling terms, of the form $L_m\tilde{R}^2$  This action can be reformulated in the ordinary Riemann geometry, by taking into account the straightforward relation between the Weyl and Ricci scalars. Moreover, the gauge condition on the Weyl vector is also imposed, and thus we obtain a gravitational action, defined in the ordinary Riemann geometry, which
contains the square of the Ricci scalar, the coupling between the Ricci
scalar and the square of the Weyl vector, some other contributions from the
Weyl geometry, as well as the geometry-matter coupling term. The field equations for this model are derived by using both
metric and Palatini formalisms.

\subsection{Basics of Weyl geometry}

We begin our investigation of the Palatini formulation of the quadratic Weyl
gravity in the presence of geometry-matter coupling by presenting first some basic elements of Weyl geometry. In the
(pseudo)-Riemannian space, in which Einstein's gravitational
field equations are formulated, the metric tensor $g_{\mu \nu}$ satisfies the metricity
condition $\nabla_{\mu} g_{\alpha \beta}=0$. From this condition one can
immediately obtain  the Levi-Civita connection, as given by
\begin{equation}  \label{eom1}
\Gamma_{\mu\nu}^\rho (g)= \frac{1}{2} g^{\rho\beta}\,(\partial_\nu
g_{\beta\mu}+\partial_\mu g_{\beta\nu}-\partial_\beta g_{\mu\nu}).
\end{equation}

By taking $\nu=\rho$ and summing over, we obtain $\Gamma_\mu\equiv
\Gamma_{\mu\nu}^\nu=\partial_\mu \ln \sqrt{-g}$, where $-g$ is the
square root of the determinant of the metric tensor $g$. Weyl conformal geometry, as well as the
corresponding gravity theory, is characterized by the presence of a vectorial
non-metricity, which implies that the covariant divergence of the metric tensor
does not vanish, like in the Riemann geometry. Hence, the basic properties of the Weyl geometry follow from the non-metricity condition
\begin{equation}  \label{1}
\tilde{\nabla}_{\lambda} g_{\mu\nu}=-\omega_{\lambda} g_{\mu\nu},
\end{equation}
where $\omega _{\lambda}$ is the Weyl vector field. From this definition we
obtain immediately $\omega_{\lambda}=(-1/4) g^{\mu\nu}\tilde{\nabla}%
_{\lambda} g_{\mu\nu}$. By using the definition of $\tilde\nabla_\mu$ in the
Weyl connection $\tilde \Gamma_{\mu\nu}^\rho$, we obtain
\begin{equation}
\tilde{\nabla}_{\lambda} g_{\mu\nu}= \partial_{\lambda} g_{\mu\nu} - \tilde{%
\Gamma}_{\mu\lambda}^\rho g_{\rho\nu}-\tilde{\Gamma}_{\nu\lambda}^\rho
g_{\mu\rho}.
\end{equation}
After performing a cyclic permutations of the indices we find
\begin{equation}
\tilde{\Gamma}_{\mu\nu}^\rho=\Gamma_{\mu\nu}^\rho(g)+ \frac{1}{2}\,
g^{\rho\lambda}\,(\tilde\nabla_\lambda g_{\mu\nu}-\tilde\nabla_\mu
g_{\nu\lambda} - \tilde\nabla_\nu g_{\lambda\mu}).
\end{equation}

By taking into account Eq.~(\ref{1}) we obtain the Weyl connection
\begin{equation}  \label{2}
\tilde\Gamma_{\mu\nu}^\rho= \Gamma_{\mu\nu}^\rho(g)+\frac{1}{2}\Big[ %
\omega_{\nu}\delta_\mu^\rho+\omega_{\mu}\delta_\nu^\rho-g_{\mu\nu}\omega^\rho%
\Big].
\end{equation}

The Weyl connection $\tilde {\Gamma}_{\mu\nu}^\rho$ is symmetric, with the
property $\tilde {\Gamma}_{\mu\nu}^\rho=\tilde {\Gamma}_{\nu\mu}^\rho$.
Hence, it follows that the standard Weyl geometry is {\it torsionless}. Additionally, $\tilde{\Gamma}$
is invariant with respect to the Weyl local gauge transformation $\Omega(x)$
of the metric $g_{\mu\nu}$
\begin{equation}  \label{3}
\hat g_{\mu\nu}=\Omega^2(x) g_{\mu\nu},\quad  \sqrt{-\hat g}=\Omega^4\sqrt{-g}.
\end{equation}

With respect to these geometric transformations, the Weyl gauge field $%
\omega_\mu$ transforms according to the rule
\begin{equation}\label{4}
\hat \omega_\mu=\omega_\mu-\partial_\mu \ln \Omega^2.
\end{equation}

Eqs.~(\ref{3}) and (\ref{4}) define a local gauged scale transformation. Using the
relation $g^{\alpha\beta}\tilde{\nabla}_\lambda g_{\alpha\beta}=2\tilde{%
\nabla}_\lambda\ln \sqrt{-g}$, for the Weyl vector field we obtain the simple expression
\begin{equation}
\omega_{\lambda}=-\frac{1}{2}\,\,\tilde{\nabla}_\lambda \ln\sqrt{-g}.
\end{equation}

We take now $\nu=\rho$ in Eq.~(\ref{2}), and, after summing over, we find
\begin{equation}
\tilde{\Gamma}_\mu=\Gamma_\mu(g)+2 \omega_{\mu}.
\end{equation}

The Riemann and Ricci tensors in Weyl geometry, as well as the Weyl scalar, are defined similarly to
their definition in Riemannian geometry, but with the replacement of the
Levi-Civita connection $\Gamma_{\mu\nu}^\rho(g)$ by the Weyl connection $%
\tilde {\Gamma}_{\mu\nu}^\rho$. Thus, we obtain
\begin{equation}
\tilde{R}^\lambda_{\,\mu\nu\sigma}(\tilde{\Gamma},g)= \partial_\nu \tilde{\Gamma}%
^\lambda_{\mu\sigma} -\partial_\sigma\tilde{\Gamma}^\lambda_{\mu\nu} +
\tilde{\Gamma}^\lambda_{\nu\rho}\,\tilde{\Gamma}^\rho_{\mu\sigma}
-\tilde\Gamma^\lambda_{\sigma\rho}\,\tilde{\Gamma}^\rho_{\mu\nu},
\end{equation}
and
\begin{equation}
\tilde{R}_{\mu\sigma}(\tilde{\Gamma},g)=\tilde{R}^\lambda_{\,\,\mu\lambda\sigma}(\tilde{%
\Gamma},g), \tilde{R}(\tilde{\Gamma},g)=g^{\mu\sigma}\tilde{R}_{\mu\sigma}(\tilde{%
\Gamma},g),
\end{equation}

Since $\tilde{\Gamma}$ is invariant under the gauge transformations (\ref{3}) and (\ref%
{4}), it follows that the Riemann and Ricci tensors of the Weyl geometry
are also scale invariant. Due to the presence of $g^{\mu\nu}$ in the Weyl
scalar curvature $\tilde{R}(\tilde{\Gamma},g)$, the Weyl scalar also transforms covariantly according to
\begin{equation}
\hat{R}(\tilde{\Gamma},g)=\frac{1}{\Omega^2}\tilde{R}(\tilde{\Gamma},g).
\end{equation}

Using the expression of $\tilde{\Gamma}$, we finally obtain
\begin{equation}  \label{R}
\tilde{R}(\tilde{\Gamma},g) =R(\Gamma,g) - 3 \nabla_{\mu} \omega^{\mu} -%
\frac{3}{2} g^{\mu\nu}\omega_{\mu} \omega_{\nu},
\end{equation}
where $R(\Gamma,g)$ is the Riemann scalar curvature, while $%
\nabla_{\mu}\omega ^\mu$ is defined by the Levi-Civita connection.

\subsection{Conformally invariant Weyl gravity}

Before beginning the in-depth investigation of the Palatini formulation of the conformally invariant Weyl gravity, we need to mention first that the Ricci tensor $\tilde{R}_{[\mu \nu}]$ has an antisymmetric component $\tilde{R}_{[\mu \nu]} \equiv (1/2)\left(R_{\mu \nu}-R_{\nu \mu}\right)$, given explicitly by
\be
\tilde{R}_{[\mu \nu]}=\frac{1}{2}\left(\partial _{\mu}\tilde{\Gamma}_{\rho \nu}^{\rho}-\partial _{\nu}\tilde{\Gamma}_{\rho \mu}^{\rho}\right).
\ee

If one introduces the Weyl gauge field $\omega _{\mu}=(1/2)\left(\tilde{\Gamma}_{\mu}-\Gamma _{\mu}\right)$, then one can show easily that $\tilde{R}_{[\mu \nu]}$ takes the form of the strength tensor of a Maxwell type field,
\be\label{F}
\tilde{R}_{[\mu \nu]} =\partial _{\mu}\omega _{\nu}-\partial _{\nu}\omega _{\mu}\equiv \tilde{F}_{\mu \nu}(\omega).
\ee

In the following we will consider the simplest version of the Weyl conformal gravity {\it in the presence of matter},  with action given by
\begin{align}
S_{W}\left(\tilde{\Gamma},g,L_m\right)&=\int \Big[\frac{1}{4!}\frac{1}{\xi ^{2}}%
\tilde{R}^{2}\left(\tilde{\Gamma},g\right)-\frac{1}{4}\,\tilde{R}_{[\mu \nu]
}^{2}\nonumber\\
&-\frac{1}{4!\gamma ^2}L_m\tilde{R}^{2}\left(\tilde{\Gamma},g\right)\Big]\sqrt{-g}d^{4}x,
\end{align}
where $\xi$ and $\gamma$ are two coupling constants.
By taking into account Eq.~(\ref{F}) for $R_{[\mu \nu]}$, we obtain the following action for the conformally invariant Weyl type gravitational theory in the presence of matter,
\begin{align}\label{S}
S_{W}\left(\tilde{\Gamma},g,L_m,\omega\right)&=\int \Big[\frac{1}{4!}\frac{1}{\xi ^{2}}\left(1-\frac{\xi ^2}{\gamma ^2}L_m\right)%
\tilde{R}^{2}\left(\tilde{\Gamma},g\right)\nonumber\\
&-\frac{1}{4}\,\tilde{F}_{\mu \nu
}^{2}(\omega)\Big]\sqrt{-g}d^{4}x.
\end{align}

The action (\ref{S}) \textit{is defined in Weyl geometry}, with the field
strength tensor $\tilde{F}_{\mu \nu }$ given by as $\tilde{F}_{\mu \nu }=%
\tilde{\nabla}_{\mu }\omega _{\nu }-\tilde{\nabla}_{\nu }\omega _{\mu
}=\nabla _{\mu }\omega _{\nu }-\nabla _{\nu }\omega _{\mu }$.

After substituting $\tilde{R}\left( \tilde{\Gamma},g\right) $ with its Riemannian
counterpart obtained from Eq.~(\ref{R}), we find the corresponding action in
Riemann geometry as given by
\begin{eqnarray}
&&S_{R}\left( \Gamma ,g,L_m,\omega\right) =\frac{1}{4!}\frac{1}{\xi ^{2}}\int \Bigg\{\left(1-\frac{\xi ^2}{\gamma ^2}L_m\right)\Bigg[ %
R^{2}\left( \Gamma ,g\right)\nonumber\\
&&-3R\left( \Gamma ,g\right) \left( \omega
^{2}+2\nabla _{\mu }\omega ^{\mu }\right)  +9\left( \nabla _{\mu }\omega ^{\mu }\right) ^{2}+9\omega ^{2}\nabla
_{\mu }\omega ^{\mu }\nonumber\\
&&+\frac{9}{4}\omega ^{4}\Bigg]
-6\xi ^{2}\,\tilde{F}_{\mu \nu
}^{2}\Bigg\} \sqrt{-g}d^{4}x,
\end{eqnarray}
where we have denoted $\omega ^{2}=g_{\mu \nu }\omega ^{\mu }\omega ^{\nu }$.

By performing a gauge transformation of the Weyl vector field we can always
find a gauge in which $\nabla _{\mu }\omega ^{\mu }=0$. Hence, \textit{by
imposing the gauge condition on the Weyl vector}, the action in the Riemann
geometry becomes
\begin{eqnarray}\label{17}
&&S_{R}\left( \Gamma ,g,L_m,\omega\right) =\frac{1}{4!}\frac{1}{\xi ^{2}}\int\Bigg\{\left(1-\frac{\xi ^2}{\gamma ^2}L_m\right) \Bigg[ %
R^{2}\left( \Gamma ,g\right) \nonumber\\
&&-3R\left( \Gamma ,g\right) \omega ^{2}
+\frac{9}{4}\omega ^{4}\Bigg]-6\xi ^2\,\tilde{F}_{\mu \nu }^{2}\Bigg\} \sqrt{-g}%
d^{4}x.
\end{eqnarray}

For the sake of generality, and with the main goal of considering the Palatini versions of the Weyl gravitational theories constructed generally from Weyl geometry, in the following we consider a general action, \textit{not necessarily conformal invariant}, which in the equivalent \textit{Riemann
geometry} can be formulated as
\begin{eqnarray}  \label{act}
S &=&\int \Bigg\{ \frac{1}{2}f_{1}\left[ R(g,\Gamma)\right]G\left(L_m\right) +f_{2}%
\left[ R(g,\Gamma)\right]G\left(L_m\right) \omega ^{2}  \nonumber \\
&&+\frac{9}{4}G\left(L_m\right)\omega ^{4}-\frac{1}{4}\,\tilde{F}_{\mu \nu }^{2}\Bigg\} %
 \sqrt{-g}d^{4}x,
\end{eqnarray}%
where $f_{i}(R)$, $i=1,2$, and $G\left(L_m\right)$ are arbitrary functions of the Ricci scalar $%
R=g^{\mu \nu }\bar{R}_{\mu \nu }$ and of the matter Lagrangian. The only condition for the functions $%
f_{i}(R)$, $i=1,2$, and $G\left(L_m\right)$ is the requirement they are analytical functions of the
Ricci scalar $R$ and of the matter Lagrangian $L_m$, that is, they must possess a Taylor series expansion about
any point $R$ and $L_m$. {\it The conformally invariant Weyl geometric gravity is a particular case of the action} (\ref{act}).

From its mathematical definition it turns out that the Riemann tensor $\bar{R}_{\mu \nu \lambda}^{\rho}$
{\it is constructed entirely in terms of the Riemannian connection} $\bar{\Gamma}$%
. The Ricci tensor is defined according to \cite{LaLi, Rub}
\begin{equation}\label{22}
\bar{R}_{\mu \nu }=\partial _{\lambda }\bar{\Gamma}_{\mu \nu }^{\lambda
}-\partial _{\nu }\bar{\Gamma}_{\mu \lambda }^{\lambda }+\bar{\Gamma}_{\mu
\nu }^{\lambda }\bar{\Gamma}_{\lambda \alpha }^{\alpha }-\bar{\Gamma}_{\mu
\lambda }^{\alpha }\bar{\Gamma}_{\nu \alpha }^{\lambda }.
\end{equation}

 In the Palatini approach, the gravitational action is formally the
same as in the standard general relativistic case, but {\it the Riemann tensor and the Ricci tensor
are constructed with the independent connection} $\bar{\Gamma}$. The variational procedure consists in {\it varying the action
independently with respect to both the metric and to the connection},
respectively. Thus, the connection $\bar{\Gamma}_{\mu \nu }^{\lambda }$ is
obtained by varying the gravitational field action, and it is not directly
constructed from the metric by using the Levi-Civita formalism.

\subsection{Gravitational field equations}

In the following we will consider the field equations obtained from the
action (\ref{act}) in the Palatini formalism. In order to do so we need to
vary the gravitational action independently with respect to the metric and
the connection.

In the following {\it we define the Ricci scalar in terms of the two independent variables} $\left(g,\bar{\Gamma}\right)$ as
\be
R=g^{\mu \nu }\bar{R}_{\mu \nu },
\ee
with $\bar{R}_{\mu \nu }$ defined in Eq.~(\ref{22}).

We also introduce the matter energy-momentum tensor of the matter according to the definition
\begin{equation}
T_{\mu \nu }=-\frac{2}{\sqrt{-g}}\frac{\delta \left( \sqrt{-g}L_{m}\right)
}{\delta g^{\mu \nu }},
\end{equation}
thus obtaining
\begin{equation}
\frac{\delta L_{m}}{\delta g^{\mu \nu }}=-\frac{1}{2}T_{\mu \nu }+\frac{1}{%
2}L_{m}g_{\mu \nu }.
\end{equation}

\subsubsection{The metric field equations}

By taking the variation of the action (\ref{act}) with {\it respect to the
metric only}, and taking into account that $g$ and $\bar{\Gamma}$ are independent
variables, thus {\it keeping the connection constant}, we immediately obtain,
\begin{eqnarray}\label{field1}
&&\left[ \frac{1}{2}f_{1}^{\prime }(R)+f_{2}^{\prime }(R)\omega ^{2}\right]
\bar{R}_{\mu \nu }+f_{2}(R)\omega _{\mu }\omega _{\nu }+\frac{9}{2}\omega
^{2}\omega _{\mu }\omega _{\nu }\nonumber  \\
&&  -\frac12\left[\frac{1}{2}f_{1}(R)+f_{2}(R)\omega ^{2}+\frac{9}{4}\omega ^{4}\right] g_{\mu \nu }\nonumber\\
&&+\frac{1}{2}\left[\frac{1}{2}f_1(R)+f_2(R)\omega ^2+\frac{9}{4}\omega ^4\right]\frac{G'\left(L_m\right)}{G\left(L_m\right)}\left(L_mg_{\mu \nu}-T_{\mu \nu}\right)\nonumber\\
-&&
\frac{1}{2G\left(L_m\right)}\tilde{T}_{\mu \nu }^{(\omega)}=0,
\end{eqnarray}%
where by a prime we have denoted a derivative with respect to the argument, $f_{i}^{\prime }(R)=df_{i}(R)/dR$, $i=1,2$, $G'\left(L_m\right)=dG\left(L_m\right)/dL_m$, and we have denoted
\be
\tilde{T}_{\mu \nu }^{(\omega)}=\frac{1}{2\sqrt{-g}}\frac{\delta}{\delta g^{\mu\nu}}\left(\sqrt{-g}\tilde{F}_{\mu \nu}^2\right).
\ee

Alternatively, the metric field equations can be written as
\bea\label{eqFK}
&&\bar{R}_{\mu \nu }+\frac{G\left( L_{m}\right) }{F}\Big[2f_{2}(R)+9\omega ^{2}\Big]\omega _{\mu
}\omega _{\nu }\nonumber\\
&&-\frac{K}{2F}\left[ \left( 1-\frac{G^{\prime }\left(
L_{m}\right) }{G\left( L_{m}\right) }L_{m}\right) g_{\mu \nu }+%
T_{\mu \nu }\right] -\frac{1}{F}\tilde{T}_{\mu \nu }^{(\omega )}=0,\nonumber\\
\eea
where we have denoted
\be\label{F}
\hspace{-1.3cm}F\left(R,\omega ^2,L_m\right)=\left[ f_{1}^{\prime }(R)+2f_{2}^{\prime
}(R)\omega ^{2}\right] G\left( L_{m}\right) ,
\ee
and
\be
K\left(R,\omega ^2,L_m\right)=\left[
f_{1}(R)+2f_{2}(R)\omega ^{2}+\frac{9}{2}\omega ^{4}\right] G\left( L_{m}\right) ,
\ee
respectively.

\subsubsection{Variation with respect to the Weyl vector}

By taking the variation of the action (\ref{act}) with respect to the Weyl vector $\omega$ we obtain the evolution equation
\be
\nabla _{\mu}\tilde{F}^{\mu \nu}+2f_2(R)G\left(L_m\right)\omega^\nu +9G\left(L_m\right)\omega ^2\omega^\nu=0.
\ee

 In Riemann geometry, the Weyl field strength tensor $\tilde{F}^{\mu \nu }$
satisfies automatically, due to its antisymmetry properties,  the equations
\begin{equation}\label{Proca2}
\nabla _{\sigma }\tilde{F}_{\mu \nu }+\nabla _{\mu }\tilde{F}_{\nu \sigma }+\nabla
_{\nu }\tilde{F}_{\sigma \mu }=0.
\end{equation}

By using the mathematical relations $\tilde{F}^{\mu \nu}=g^{\alpha \mu}g^{\beta \nu}\tilde{F}_{\alpha \beta}=g^{\alpha \mu}g^{\beta \nu}\left(\nabla _{\alpha}\omega _{\beta}-\nabla _{\beta}\omega _{\alpha}\right)=\nabla ^{\mu}\omega ^{\nu}-\nabla ^{\nu}\omega ^{\mu}$, we immediately find
\bea
\nabla _{\mu }\tilde{F}^{\mu \nu }&=&\nabla _{\mu }\nabla ^{\mu }\omega ^{\nu }-\nabla _{\mu  }\nabla ^{\nu }\omega ^{\mu }\nonumber\\
&=&\nabla _{\mu
}\nabla ^{\mu }\omega ^{\nu }-R_{\beta }^{\nu }\omega ^{\beta}-\nabla ^{\nu }\left( \nabla _{\mu }\omega ^{\mu }\right) ,
\eea
where the definitions of the Riemann tensor \cite{LaLi},
\be
\left(\nabla _{\mu }\nabla _{\nu }-\nabla _{\nu }\nabla _{\mu }\right) A^{\alpha }=-A^{\beta }R_{~\beta \nu \mu
}^{\alpha },
\ee
and of its contraction,
\be
\left( \nabla _{\mu }\nabla _{\nu }-\nabla _{\nu }\nabla _{\mu }\right) A^{\mu }=A^{\beta }R_{\beta \nu },
\ee
where used. Therefore, it follows that the Weyl vector satisfies the generalized wave equation
\begin{align}
\Box \omega ^{\nu}-R_{\beta }^{\nu }\omega ^{\beta}+2f_2(R)G\left(L_m\right)\omega^\nu +9G\left(L_m\right)\omega ^2\omega^\nu=0,
\end{align}
where the gauge condition for $\omega _{\mu}$, $\nabla _{\mu}\omega ^{\mu}=0$, has also been used.

\subsubsection{Variation with respect to the connection}

In the Palatini formalism the next step in obtaining the field equations requires the variation of the action
with respect to the independent connection $\bar{\Gamma }$. The variation can be done by using
{\it the Palatini identity}
\begin{equation}
\delta \bar{R}_{\mu \nu }=\bar{\nabla} _{\lambda }\left( \delta \bar{%
\Gamma }_{\mu \nu }^{\lambda }\right) -\bar{\nabla }_{\mu }\left( \delta
\bar{\Gamma }_{\nu \lambda}^{\lambda }\right) ,
\end{equation}%
where $\bar{\nabla }_{\lambda }$ {\it is the covariant derivative associated
with the connection}  $\bar{\Gamma}$. In the following {\it we assume that the Weyl vector $\omega$, as well as the Weyl field strength $\tilde{F}_{\mu \nu}$ and the matter Lagrangian $L_m$, are independent on the connection} $\bar{\Gamma}$.

By taking the variation of the action (\ref{act}) with respect to the
connection $\bar{\Gamma }$ we obtain
\begin{equation}
\frac{\delta S}{\delta \bar{\Gamma }}=\frac{1}{2}\int B^{\mu \nu }\left[
\bar{\nabla } _{\lambda }\left( \delta \bar{\Gamma }_{\mu \nu }^{\lambda
}\right) -\bar{\nabla }_{\mu }\left( \delta \bar{\Gamma }_{\nu \lambda
}^{\lambda }\right) \right] \sqrt{-g}d^{4}x,
\end{equation}%
where we have denoted
\begin{equation}
B^{\mu \nu }=\left[ \frac{1}{2}f_{1}^{\prime }(R)+f_{2}^{\prime }(R)\omega ^2 \right] G\left(L_m\right)g^{\mu \nu }.
\end{equation}

By integrating by parts we immediately find
\begin{eqnarray}
\frac{\delta S}{\delta \bar{\Gamma }}&=&\frac{1}{2}\int \bar{\nabla }%
_{\lambda }\left[ \sqrt{-g}\left( B^{\mu \nu }\delta \bar{\Gamma }_{\mu
\nu }^{\lambda }-B^{\lambda \nu }\delta \bar{\Gamma }_{\nu \alpha
}^{\alpha }\right) \right] d^{4}x+  \nonumber \\
&&+\frac{1}{2}\int \bar{\nabla }_{\mu }\left[ \sqrt{-g}\left( B^{\mu \nu
}\delta _{\alpha }^{\lambda }-B^{\lambda \nu }\delta _{\alpha }^{\mu
}\right) \right] \delta \bar{\Gamma }_{\lambda \nu }^{\alpha }d^{4}x.\nonumber\\
\end{eqnarray}

In $\delta S/\delta \tilde{\Gamma }$ the first term  is a total derivative,
and thus it can be removed. Hence the variation of the action (\ref{act}) with respect
to the connection becomes
\begin{equation}\label{con}
\bar{\nabla }_{\mu }\left[ \sqrt{-g}\left( B^{\mu \lambda}\delta _{\alpha
}^{\nu }-B^{\lambda \nu }\delta _{\alpha }^{\mu }\right) \right] =0.
\end{equation}

One can further simplify Eq.~(\ref{con})  by  taking into account that
when $\alpha =\nu $,  $\bar\nabla_\mu (\sqrt{-g}B^{\mu\lambda})=0$.  Substituting back to Eq.~(\ref{con}), one obtains
\begin{equation}\label{con1}
\bar{\nabla }_{\alpha }\left\{ \sqrt{-g}\left[ f_{1}^{\prime
}(R)+2f_{2}^{\prime }(R)\omega ^2 \right]G\left(L_m\right) g^{\mu \nu }\right\}
=0,
\end{equation}
or, equivalently,
\be
\bar{\nabla }_{\alpha }\left\{ \sqrt{-g}F\left(R, \omega ^2,L_m\right) g^{\mu \nu }\right\}=0,
\ee
where $F$ is defined in Eq.~(\ref{F}).

Eq.~(\ref{con1}) shows that {\it the connection $\bar{\Gamma}$  is {\it compatible with a
conformal metric}}. We introduce now {\it a new metric}
$h_{\mu \nu}$, {\it conformal} to $g_{\mu \nu }$, and defined according to
\bea
h_{\mu \nu }&\equiv& \left[ f_{1}^{\prime }(R)+2f_{2}^{\prime }(R)\omega ^2 \right]G\left(L_m\right) g_{\mu \nu }\nonumber\\
&\equiv& F\left(R, \omega ^2,L_m\right)g_{\mu \nu}.
\eea
Hence we obtain
\bea
\sqrt{-h}h^{\mu \nu }&=&\sqrt{-g}\left[ f_{1}^{\prime }(R)+2f_{2}^{\prime
}(R)\omega ^2 \right]G\left(L_m\right) g^{\mu \nu }\nonumber\\
&=&\sqrt{-g}F\left(R, \omega ^2,L_m\right)g^{\mu \nu},
\eea
where $h$ is the determinant of the metric $h_{\mu \nu}$. Thus Eq.~(\ref%
{con1}) becomes the definition of the Levi-Civita connection $\bar{\Gamma }
$ of $h_{\mu \nu }$, giving
\begin{equation}
\bar{\Gamma }_{\mu \nu }^{\lambda }=\frac{1}{2}h^{\lambda \rho }\left(
\partial _{\nu }h_{\mu \rho }+\partial _{\mu }h_{\nu \rho }-\partial _{\rho
}h_{\mu \nu }\right) .
\end{equation}

By taking into account the explicit form of $h_{\mu \nu }$ we obtain
\begin{equation}  \label{con2}
\bar{\Gamma }_{\mu \nu }^{\lambda }=\frac{1}{2}\frac{g^{\lambda \rho }}{F}%
\left[ \partial _{\nu }\left( Fg_{\mu \rho }\right) +\partial _{\mu }\left(
Fg_{\nu \rho }\right) -\partial _{\rho }\left( Fg_{\mu \nu }\right) \right],
\end{equation}
where we have denoted $F=F\left(R,\omega ^2,L_m\right)=\left[f_{1}^{\prime }(R)+2f_{2}^{\prime }(R)\omega ^2\right]G\left(L_m\right)$.

In terms of the Levi-Civita connection $\Gamma _{\mu \nu}^{\lambda }$
associated to the metric $g$,
\begin{equation}  \label{con3}
\Gamma _{\mu \nu }^{\lambda }=\frac{1}{2}g^{\lambda \rho }\left( \partial
_{\nu }g_{\mu \rho }+\partial _{\mu }g_{\nu \rho }-\partial _{\rho }g_{\mu
\nu }\right),
\end{equation}
$\bar{\Gamma }_{\mu \nu }^{\lambda }$ can be expressed as
\begin{equation}\label{con4}
\bar{\Gamma}_{\mu \nu }^{\lambda }=\Gamma _{\mu \nu }^{\lambda }+\partial
_{\nu }\ln \sqrt{F}\delta _{\mu }^{\lambda }+\partial _{\mu }\ln \sqrt{F}%
\delta _{\nu }^{\lambda }-g_{\mu \nu }g^{\lambda \rho }\partial _{\rho }\ln
\sqrt{F}.
\end{equation}

The tensor $\bar{R}_{\mu \nu}$, constructed from the metric by using the Levi-Civita connection as defined in Eq.~(\ref{con4}), is given, in terms of the Ricci tensor $R_{\mu \nu}$,  by \cite{Pal1,Pal2},
\begin{eqnarray}\label{ricci}
\bar{R}_{\mu \nu }&=&R_{\mu \nu }(g)+\frac{3}{2}\frac{1}{F^{2}}\left(
\nabla _{\mu }F\right) \left( \nabla _{\nu }F\right)  \nonumber \\
&&-\frac{1}{F}\left( \nabla _{\mu }\nabla _{\nu }+\frac{1}{2}g_{\mu \nu }
\Box \right)F .
\end{eqnarray}%

The Ricci scalar and the Einstein tensor can be immediately obtained as
\begin{equation}\label{37}
\bar{R}=R\left( g\right)-3\frac{1}{F}\square F+\frac{3}{2}\frac{1}{F^{2}}%
\left( \nabla _{\mu }F\right) \left( \nabla ^{\mu }F\right) ,
\end{equation}
and
\begin{eqnarray}  \label{ein}
\bar{G}_{\mu \nu }&=&\bar{R}_{\mu \nu}-\frac{1}{2}g_{\mu \nu }\tilde{R}%
=G_{\mu \nu }(g)+\frac{3}{2}\frac{1}{F^{2}}\left( \nabla _{\mu }F\right)
\left( \nabla _{\nu }F\right)  \nonumber \\
&&\hspace{-1.0cm}-\frac{1}{F}\left( \nabla _{\mu }\nabla _{\nu }-g_{\mu \nu
}\square \right)F -\frac{3}{4}\frac{1}{F^{2}}g_{\mu \nu }\left( \nabla
_{\lambda }F\right) \left( \nabla ^{\lambda }F\right) ,
\end{eqnarray}
respectively, with {\it all covariant derivatives taken with respect to the
metric $g_{\mu \nu}$}.

\subsubsection{Gravitational field equations in the Palatini formalism}

By using the expression of the Ricci tensor given by Eq.~(\ref{37}%
), the gravitational field equation Eq.~(\ref{eqFK}) can be written as
\begin{eqnarray}\label{fieldp}
&&\bar{G}_{\mu \nu }+\frac{1}{2}\left[ R-3\frac{1}{F}\square F+\frac{3}{2}%
\frac{1}{F^{2}}\left( \nabla _{\mu }F\right) \left( \nabla ^{\mu }F\right) %
\right] g_{\mu \nu }+ \nonumber\\
&&+\frac{G\left( L_{m}\right) }{F}\Big(2f_{2}(R)+9\omega ^{2}\Big)\omega _{\mu
}\omega _{\nu }\nonumber\\
&&-\frac{K}{2F}\left[ \left( 1-\frac{G^{\prime }\left(
	L_{m}\right) }{G\left( L_{m}\right) }L_{m}\right) g_{\mu \nu }+%
T_{\mu \nu }\right] -\frac{1}{F}\tilde{T}_{\mu \nu }^{(\omega )}=0.\nonumber\\
\end{eqnarray}

By substituting the expression of the Einstein tensor as given by Eq.~(\ref{ein}%
) into the field equation Eq.~(\ref{fieldp}), we obtain the gravitational
field equation of the Weyl geometric gravity theory in the presence of a nonminimal coupling between matter
and geometry in the Palatini formalism as
\begin{eqnarray}\label{fieldf}
&&G_{\mu \nu }+\frac{3}{2}\frac{1}{F^{2}}\left( \nabla _{\mu }F\right)
\left( \nabla _{\nu }F\right) -\frac{1}{F}\left( \nabla _{\mu }\nabla _{\nu
}\right) F-\frac{1}{2F}g_{\mu \nu }\square F\nonumber\\
&&+\frac{1}{2}Rg_{\mu \nu }+\frac{G\left( L_{m}\right) }{F}\Big[2f_{2}(R)+9\omega ^{2}\Big]\omega _{\mu
}\omega _{\nu }\nonumber\\
&&-\frac{K}{2F}\left[ \left( 1-\frac{G^{\prime }\left(
	L_{m}\right) }{G\left( L_{m}\right) }L_{m}\right) g_{\mu \nu }+%
\frac{G^{\prime }\left(
	L_{m}\right) }{G\left( L_{m}\right) }T_{\mu \nu }\right] \nonumber\\
&&-\frac{1}{F}\tilde{T}_{\mu \nu }^{(\omega )}=0.
\end{eqnarray}

Taking the trace of the metric field equation Eq.~(\ref{eqFK}) we obtain
\bea\label{trace}
&&\bar{R}+\frac{G\left( L_{m}\right) }{F}\Big(2f_{2}(R)+9\omega ^{2}\Big)\omega ^2\nonumber\\
&&-\frac{K}{2F}\left[4 \left( 1-\frac{G^{\prime }\left(
	L_{m}\right) }{G\left( L_{m}\right) }L_{m}\right)+%
\frac{G^{\prime }\left(
	L_{m}\right) }{G\left( L_{m}\right) }T\right]=0,
\eea
where the trace $T_{\mu}^{(\omega )\;\mu}$ of the energy-momentum tensor of the Weyl field identically vanishes, $T_{\mu}^{(\omega )\;\mu}=0$.

By using  Eq.~(\ref{37}), we find the equation determining $R$ as a
function of $\omega$ as
\begin{align}
R\left( g\right)&-3\frac{1}{F}\square F+\frac{3}{2}\frac{1}{F^{2}}%
\left( \nabla _{\mu }F\right) \left( \nabla ^{\mu }F\right)\nonumber\\&+\frac{G\left( L_{m}\right) }{F}\Big(2f_{2}(R)+9\omega ^{2}\Big)\omega ^2\nonumber\\
&-\frac{K}{2F}\left[4 \left( 1-\frac{G^{\prime }\left(
	L_{m}\right) }{G\left( L_{m}\right) }L_{m}\right)+%
\frac{G^{\prime }\left(
	L_{m}\right) }{G\left( L_{m}\right) }T\right]=0.
\end{align}
Also, the covariant divergence of the energy-momentum tensor becomes
\begin{widetext}
\begin{align}\label{57}
\nabla^\alpha T_{\alpha\nu}&=\frac{4G^2}{G^\prime K}\left(f_2+\frac92\omega^2\right)\left(\omega^\alpha\nabla_\nu\omega_\alpha+\omega_\nu\nabla_\alpha\omega^\alpha-\omega_\nu\nabla^\alpha\ln F+\omega^\alpha\omega_\nu\frac{\nabla_\alpha L_m}{G}\right)+(T_{\alpha\nu}-g_{\alpha\nu}L_m)\nabla^\alpha\ln\left(\frac{FG}{G^\prime K}\right)\nonumber\\
&-\frac{2G}{G^\prime K}\left[(FR_{\alpha\nu}-T_{\alpha\nu})-\frac12g_{\mu\nu}\left(K+\frac{4F\Box F-6\nabla^\alpha F\nabla_\alpha F}{F}\right)\right]\nabla^\alpha\ln F+\frac{4G^2\omega^\alpha\omega_\nu}{G^\prime K}\nabla_\alpha f_2+\nabla_\nu L_m\nonumber\\
&+\frac{G}{FG^\prime K}(5\nabla^\alpha F\nabla_\nu\nabla_\alpha F-3F\nabla_\nu\Box F+F^2\nabla_\nu R-F\nabla_\nu K)+\frac{36G^2\omega^\alpha\omega^\mu}{G^\prime K}(\omega_\nu\nabla_\mu\omega_\alpha+\omega_\alpha F_{\mu\nu})
\end{align}
\end{widetext}

As one can see from Eq.~(\ref{57}), the covariant divergence of the matter energy-momentum tensor is not conserved in the Weyl geometric formulation of $f\left(R,L_m\right)$ theory. This result is similar to the one obtained in the metric case \cite{HaSh}, and it is essentially a direct consequence of the presence of the geometry-matter coupling. From a physical point of view, this result can be interpreted as describing irreversible particle production, and an energy transfer from gravity (geometry) to matter \cite{Ha9}. Such creation processes may play an important role in cosmology, and may also provide a mechanism that may explain the late acceleration of the Universe \cite{Wu}.

\section{Palatini formulation of quadratic Weyl gravity with matter-curvature coupling}\label{sect2}

The general formalism developed in the previous Sections can be immediately applied to the case of conformally invariant actions. In the particular case of the simplest conformally invariant Weyl action (\ref{17}), with
\be
f_1(R)=\frac{1}{12\xi^2}R^2\left(g,\bar{\Gamma}\right),
\ee
\be
f_2(R)=-\frac{1}{8\xi^2}R\left(g,\bar{\Gamma}\right),
\ee
and
\be
G\left(L_m\right)=1-\frac{\xi^2}{\gamma ^2}L_m,
\ee
respectively. Then we obtain immediately
\be
F\left(R,\omega ^2,L_m\right)=\frac{1}{2\xi^2}\left(\frac{R}{3}-\frac{\omega ^2}{2}\right)\left(1-\frac{\xi^2}{\gamma ^2}L_m\right),
\ee
and
\bea
K\left(R,\omega ^2,L_m\right)&=&\left(\frac{1}{12\xi^2}R^2-\frac{1}{4\xi^2}R\omega^2+\frac{9}{2}\omega ^4\right)\nonumber\\
&&\times \left(1-\frac{\xi ^2}{\gamma^2}L_m\right),
\eea
respectively. The conformally related metric $h_{\mu \nu}$ is given by
\be
h_{\mu \nu}=\frac{1}{2\xi^2}\left(\frac{R}{3}-\frac{\omega ^2}{2}\right)\left(1-\frac{\xi^2}{\gamma ^2}L_m\right)g_{\mu \nu}.
\ee

\subsection{Palatini formulation of quadratic Weyl gravity in vacuum}

If the neglect the effect of the matter by taking $L_m=0$,  then $G\left(L_m\right)=1$, and
\be
\hspace{-1.2cm}F\left(R,\omega ^2\right)=\frac{1}{6\xi^2}\left(R-\frac32\omega ^2\right),
\ee
and
\be
K\left(R,\omega ^2\right)=\frac{1}{4\xi ^2}\left(\frac{R^2}{3}-R\omega ^2\right)+\frac{9}{2}\omega ^4,
\ee
respectively.  The metric field equations of the Palatini formalism of the quadratic Weyl geometric gravity take the form
\bea\label{42}
&&\left(R-\frac32\omega ^2\right)\bar{R}_{\mu \nu}-\frac32(R-36\omega^2)\omega _{\mu}\omega _{\nu}\nonumber\\
&&-\frac14\left(R^2-3\omega ^2R+54\omega ^4\right)g_{\mu \nu}-6\xi ^2 \tilde{T}_{\mu \nu}^{(\omega)}=0.
\eea

The metric $h_{\mu \nu}$ conformal to $g_{\mu \nu}$ is obtained as
\be
h_{\mu \nu}=\frac{1}{6\xi^2}\left(R-\frac32\omega ^2\right)g_{\mu \nu},
\ee
and it depends on the Ricci scalar as well as of the Weyl vector. The Palatini field equation (\ref{fieldf}) can be written down straightforwardly, as well as the scalar equation (\ref{42}) relating $R$ and $\omega$, and they are given by
\begin{eqnarray}
&&G_{\mu \nu }+\frac{3}{2}\frac{1}{F^{2}}\left( \nabla _{\mu }F\right)
\left( \nabla _{\nu }F\right) -\frac{1}{F}\left( \nabla _{\mu }\nabla _{\nu
}\right) F-\frac{1}{2F}g_{\mu \nu }\square F  \notag  \label{fieldf} \\
&&+\frac{1}{2}Rg_{\mu \nu }+\frac{1}{F}\Big[9\omega ^{2}-\frac{1}{4\xi ^{2}}R%
\Big]\omega _{\mu }\omega _{\nu }  -\frac{K}{2F}g_{\mu \nu }-\frac{1}{F}\tilde{T}_{\mu \nu }^{(\omega )}=0,\nonumber\\
\end{eqnarray}
and
\begin{eqnarray}
 &&R-3\frac{1}{F}\square F+\frac{3}{2}\frac{1}{F^{2}}\left( \nabla _{\mu
}F\right) \left( \nabla ^{\mu }F\right)
 +\frac{1}{F}\Big(9\omega ^{2}-\frac{1}{4\xi ^{2}}R\Big)\omega ^{2}\nonumber\\
 &&-\frac{2K}{F}=0,
\end{eqnarray}
respectively.

\subsection{Palatini formulation of the linear/scalar representation of Weyl geometric gravity}

An alternative, and equivalent description of the dynamical properties of gravitational theories based on the action (\ref{S}) was considered in \cite{Gh7} (see also references therein), and it is based on the introduction of an auxiliary scalar field $\phi _{0}$, according to the definition,
\begin{equation}\label{R2}
\tilde{R}^{2}+2\phi _{0}^{2}\tilde{R}+\phi _{0}^{4}=0.
\end{equation}
After substituting  $\tilde{R}^{2}\rightarrow-2\phi _{0}^{2}\tilde{R}-\phi _{0}^{4}$ into the action \ref{S},
and performing its variation with respect to $\phi _{0}$ we obtain
the equation
\begin{equation}
\phi _{0}\left( \tilde{R}+\phi _{0}^{2}\right) =0,
\end{equation}%
which gives for $\phi _{0}^{2}$ the expression
\begin{equation}
\phi _{0}^{2}=-\tilde{R}.
\end{equation}%
Therefore, through this substitution, we reobtain the original form of the
Lagrangian as introduced in the initial Weyl geometry. By substituting Eq. (\ref{R2}) into the action (\ref{S}), with
the use of Eq.~(\ref{R}), we obtain
\begin{eqnarray}\label{pW}
\hspace{-0.6cm}&&S_R\left(g,\bar{\Gamma}\right) =-\int \Bigg\{\frac{1}{2\xi ^{2}} \Bigg[\frac{\phi _{0}^{2}}{6}R\left(g,\bar{\Gamma}\right)\nonumber\\
\hspace{-0.6cm}&&-\frac{1}{2}%
\phi _{0}^{2}\nabla _{\mu }\omega ^{\mu }
-\frac{1}{4}\phi _{0}^{2}\omega _{\mu }\omega
^{\mu }+\frac{\phi _{0}^{4}}{12}\Bigg]+\frac{1}{4}\,\tilde{F}_{\mu \nu }^{2}\Bigg\}\sqrt{-g}%
d^{4}x.
\end{eqnarray}

 The action~(\ref{pW}) is a particular case of the general action (\ref{act}), corresponding to $f_1(R)=0$, and $f_2(R)=R$, respectively. Hence, all the previous results obtained for the Palatini version of the field equations derived for (\ref{act}) can now be applied to the linear/scalar version of quadratic Weyl gravity, by taking into account the particular forms of the functions $f_1$ and $f_2$. The Palatini formulation of the theory for this case was extensively investigated in \cite{Gh5} and \cite{Cai}, respectively, where it was shown that the basic results obtained in the metric case remain also valid in the Palatini formulation of the theory. Moreover, in the presence of non-minimally coupled scalar field (Higgs-like) with Palatini connection, the theory gives successful inflation, and a specific prediction for the tensor-to-scalar ratio $0.007 \leq r \leq 0.01$ for the current spectral index $n_s$ (at 95 \% CL) and $N =60$ efolds. The obtained value of $r$ is slightly larger than the value corresponding to inflation in Weyl quadratic gravity of similar symmetry, due to the presence of different forms of the non-metricity. Hence, one can establish a relation between non-metricity and inflationary predictions that enables the testing of the theory by using future CMB observations.

 \section{Cosmological applications}\label{sect3}

 We will investigate now the cosmological applications of the Palatini formulation of the conformally invariant $f\left(R,L_m\right)$ theory. In the following we consider a homogeneous and isotropic Universe, described by a flat FLRW space-time, with line element given by
 \begin{equation}
 ds^2=-dt^2+a(t)^2(dx^2+dy^2+dz^2),
 \end{equation}
 where $a(t)$ is the scalar factor. Moreover, we at this moment we introduce the Hubble function, defined according to $H=\dot{a}/a$. We assume that the matter content of the Universe can be described as a perfect fluid, with Lagrangian $L_m=-\rho$, where $\rho$ is the matter energy-density, with the corresponding energy-momentum tensor, defined in a comoving frame, given by
 \begin{equation}
 T^\mu_{\nu}={\rm diag}(-\rho,p,p,p).
 \end{equation}

 We also assume that the Weyl vector in the FLRW universe has only a temporal component $A_0$, and therefore it is given by
 \begin{align}
 A_\mu=(A_0,0,0,0).
 \end{align}

 This choice is suitable for cosmological applications, since it maintains the isotropy and homogeneity of the space-time.  The Friedmann and Raychaudhuri equations of the Palatini formulation of the Weyl geometric $f\left(R,L_m\right)$ gravity theory can be obtained straightforwardly from Eq.~(\ref{fieldf}), and they are given by
 \begin{align}\label{Fr1}
 3H^2&=\frac12A_0^2(9A_0^2-2f_2)\frac{G}{F}+\frac12\frac{K}{F}\nonumber\\&-3H\frac{\dot{F}}{F}-\frac34\left(\frac{\dot{F}}{F}\right)^2+\frac34\frac{K}{F}\frac{G^\prime}{G}(\rho+p),
 \end{align}
 and
  \begin{align}\label{Fr2}
 \dot{H}&=-\frac12A_0^2(9A_0^2-2f_2)\frac{G}{F}+\frac12H\frac{\dot{F}}{F}\nonumber\\&+\frac34\left(\frac{\dot{F}}{F}\right)^2-\frac12\frac{\ddot{F}}{F}-\frac14\frac{K}{F}\frac{G^\prime}{G}(\rho+p),
 \end{align}
respectively.  Also, the equation of motion of the Weyl vector can be written as
 \begin{align}\label{Fr3}
 \ddot{A}_0+3\dot{H}A_0+3H\dot{A}_0+G(9A_0^2-2f_2)=0.
 \end{align}

 In the following, instead of the time variable $t$, we use the redshift coordinate $z$, defined as
 \begin{align}
 1+z=\frac1a,
 \end{align}
 and define the dimensionless Hubble function according to
 \begin{align}
 h=\frac{H}{H_0},
 \end{align}
 where $H_0$ is the current value of the Hubble parameter $H=\dot{a}/a$.

 To describe the decelerating/accelerating nature of the cosmological evolution we use the deceleration parameter $q$, defined in the redshift space as
 \begin{align}
 q=-1+\frac{(1+z)}{h(z)}\frac{d h(z)}{dz}.
 \end{align}

\subsection{Simple cosmological toy models}

 We will consider first some simple solutions of the cosmological system of evolution equations Eqs~(\ref{Fr1})-(\ref{Fr3}).
 \begin{figure*}
 	\includegraphics[scale=0.95]{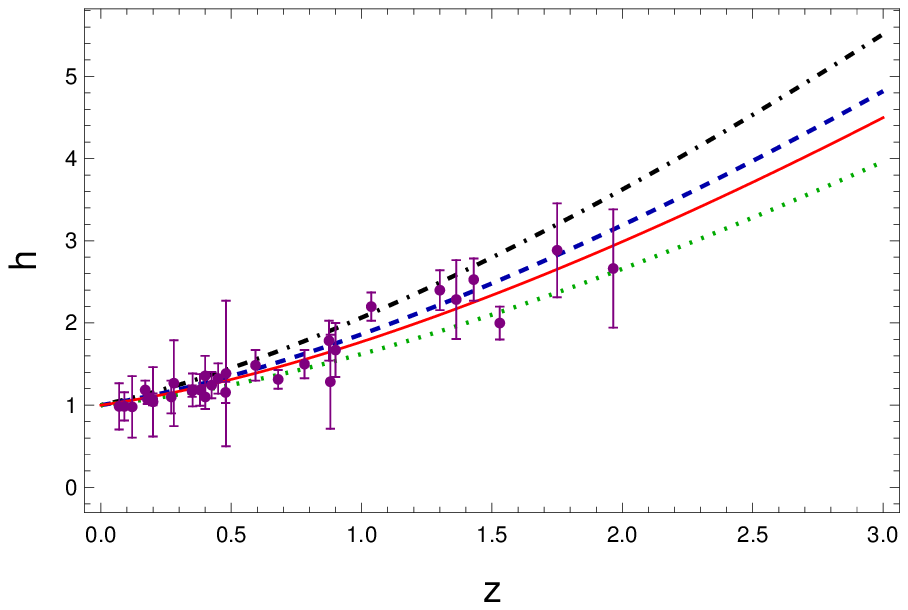}\quad\includegraphics[scale=0.95]{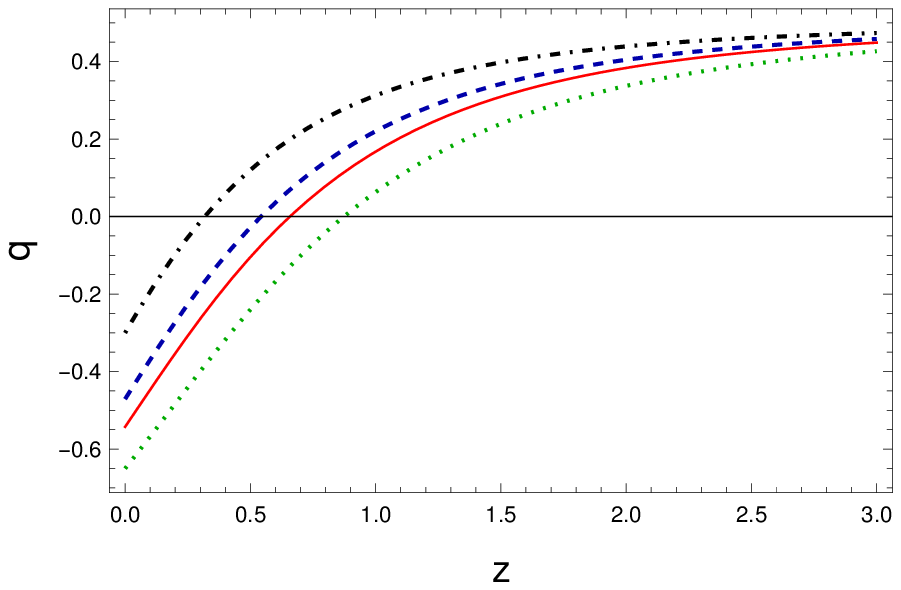}
 	\caption{\label{fig2}The Hubble function $h(z)$ and the deceleration parameters $q(z)$ of the model (\ref{func1}), corresponding to the best fit values of $\epsilon$ (dashed), $\epsilon=3.2$ (dot-dashed) and $\epsilon=4.6$ (dotted). The solid line corresponds to the $\Lambda$CDM model, and the error bars indicate the observational data with their errors.}
 \end{figure*}

As a first example,  we adopt for the functions $f_1$, $f_2$ and $G$ the following simple forms
 \begin{equation}\label{func1}
f_1=\alpha,\quad f_2=\beta R,\quad G=\gamma,
 \end{equation}
 where $\alpha$, $\beta$ and $\gamma$ are constants.

 \subsubsection{de Sitter solution}

 The cosmological model described by the functions (\ref{func1}) has an exact solution, corresponding to constant Weyl vector, given by
 \begin{align}
 A_0=\sqrt{\frac{2}{3}}\alpha^{\frac{1}{4}},\quad H_0=\frac{1}{2\sqrt{\beta}}\alpha^{\frac14}.
 \end{align}

 As one can see from the above solution, $\alpha$ and $\beta$ should be constant, but $\gamma$ remains arbitrary.

 \subsubsection{Models with constant $A_0$}

 By assuming that the temporal component of the Weyl vector is a constant, the model (\ref{func1}) has a non-trivial solution, corresponding to
 \begin{align}
 A_0=\sqrt{\frac23}\alpha^{\frac14},\qquad \gamma=-\frac{3}{4\beta},
 \end{align}
 and with the dimensionless Hubble function given by
 \begin{align}
 h(z)=\sqrt{\frac{1}{6}\left[(6-\epsilon)(1+z)^3+\epsilon\right]},
 \end{align}
respectively, where we have denoted
\be
\epsilon=\frac{3\sqrt\alpha}{3\beta H_0^2}.
\ee

 In order to find the best fit value of the parameter $\epsilon$, we use the Likelihood analysis using the observational data on the Hubble parameter in the redshift range $z\in(0,2)$.  In the case of independent data points, the likelihood function can be defined as
 \begin{align}
 L=L_0e^{-\chi^2/2},
 \end{align}
 where $L_0$ is the normalization constant and the quantity $\chi^2$ is defined as
 \begin{align}
 \chi^2=\sum_i\left(\frac{O_i-T_i}{\sigma_i}\right)^2.
 \end{align}

 Here $i$ counts the data points, $O_i$ are the observational value, $T_i$ are the theoretical values, and $\sigma_i$ are the errors associated with the $i$th data obtained from observations. By maximizing the likelihood function,  the best fit values of the parameters $\epsilon$ and $H_0$ at $1\sigma$ confidence level, can be obtained as
 \begin{align}
 \epsilon&=3.883^{+0.345}_{-0.417},\nonumber\\
 H_0&=67.05^{+2.945}_{-3.021}.
 \end{align}

\begin{figure}
 	\includegraphics[scale=0.6]{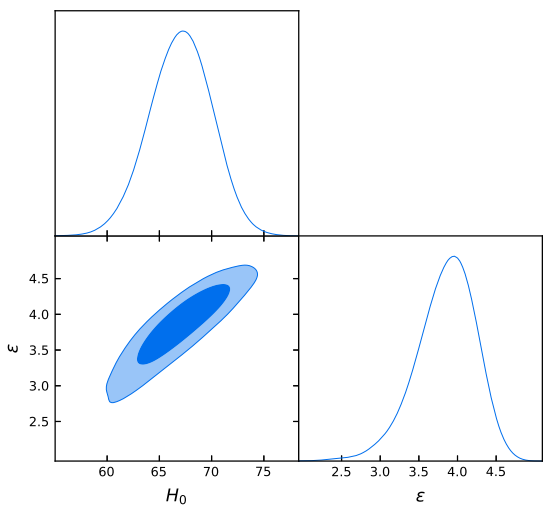}
 	\caption{\label{fig1}The corner plot for the toy model (\ref{func1}).}
 \end{figure}

 The redshift evolution of the Hubble function and of the deceleration parameter $q$ are represented, for this model, in Fig.~\ref{fig2}. As one can see from the Figure, despite its simplicity, the present model can give an acceptable description of the observational data, and at low redshifts it also reproduces the predictions of the standard $\Lambda$CDM model for the behavior of the Hubble function. However, some differences do appear in the evolution of the deceleration parameter, which may be able to provide, once the quality of the observational data improves, further cosmological tests of the Palatini formulation of the $f\left(R,L_m\right)$ theory.
 In Fig.~\ref{fig1} we have shown the corner plot corresponding to the above estimation of $\epsilon$ and $H_0$ which shows the best fit values of the model parameter together with their $1\sigma$ and $2\sigma$ confidence intervals.

\subsection{Conformally invariant models}

Let us now consider a quadratic conformally invariant Weyl geometric cosmological model, obtained by assuming for the functions $f_1$, $f_2$ and $G$ the forms
\begin{align}\label{func2}
f_1=\alpha R^2,\quad f_2=\beta,\quad G=1-\gamma L_m.
\end{align}

For the sake of simplicity, we will assume that the matter Lagrangian has the form $L_m=p$. For a dust dominated universe, the Raychaudhuri and the Weyl vector field equations can be simplified to
\begin{align}
8\beta A_0^2-9A_0^4+48\alpha\left[3\dot{H}(6H^2+\dot{H}+9H\ddot{H}+\dddot{H})\right]=0,
\end{align}
and
\begin{align}
\ddot{A}_0+3(H\dot{A}_0+A_0\dot{H})+9A_0^3-2\beta A_0=0,
\end{align}
respectively. The Friedmann equation is algebraic with respect to the energy density, and it could be used to obtain $\rho$ in terms of the Hubble function $H$, and $A_0$. Now, let us define the set of dimensionless variables
\begin{align}
&t=H_0\tau,\quad H=H_0 h,\quad A_0=H_0\bar{A}_0\nonumber\\ &\bar\rho_m=\frac{\rho_m}{6\kappa^2H_0^2},\quad \bar\gamma=6\kappa^2H_0^2\gamma,\quad \bar\beta=\frac{\beta}{H_0^2},
\end{align}
where $H_0$ is the current value of the Hubble parameter, and $\kappa^2=1/16\pi G$.  Transforming to the redshift coordinate $z$
one can solve numerically the equations \eqref{eq1} and \eqref{eq2} to obtain the evolution of $h$ and $\bar{A}_0$.

In order to find the best fit value of the parameter $\epsilon$, we use the Likelihood analysis using the observational data on the Hubble parameter in the redshift $z\in(0,2)$ \cite{hubble}. The best fit values of the model parameters $\alpha$ and $\bar\beta$, and of the Hubble parameter $H_0$ are summarized in Table~ \ref{tab1a}.
\begin{table}
		\begin{tabular}{|c||c||c||c|}
		\hline
	Parameter&Best fit&$1\sigma$ C.I.&$2\sigma$ C.I.\\
	\hline\hline
	$\alpha$&-0.73&$-0.73\pm0.02$&$-0.73\pm0.04$\\
	\hline
	$\bar\beta$&$-1.22$&$-1.22\pm0.04$&$-1.22\pm0.08$\\
	\hline
	$H_0$&$67.64$&$67.64\pm1.41$&$67.64\pm2.77$\\
	\hline
\end{tabular}
\caption{The best fit values of the parameters $\alpha$, $\bar\beta$, and $H_0$ for the conformally invariant quadratic Weyl geometric model (\ref{func2}).}\label{tab1a}
\end{table}
\begin{figure*}
	\includegraphics[scale=0.7]{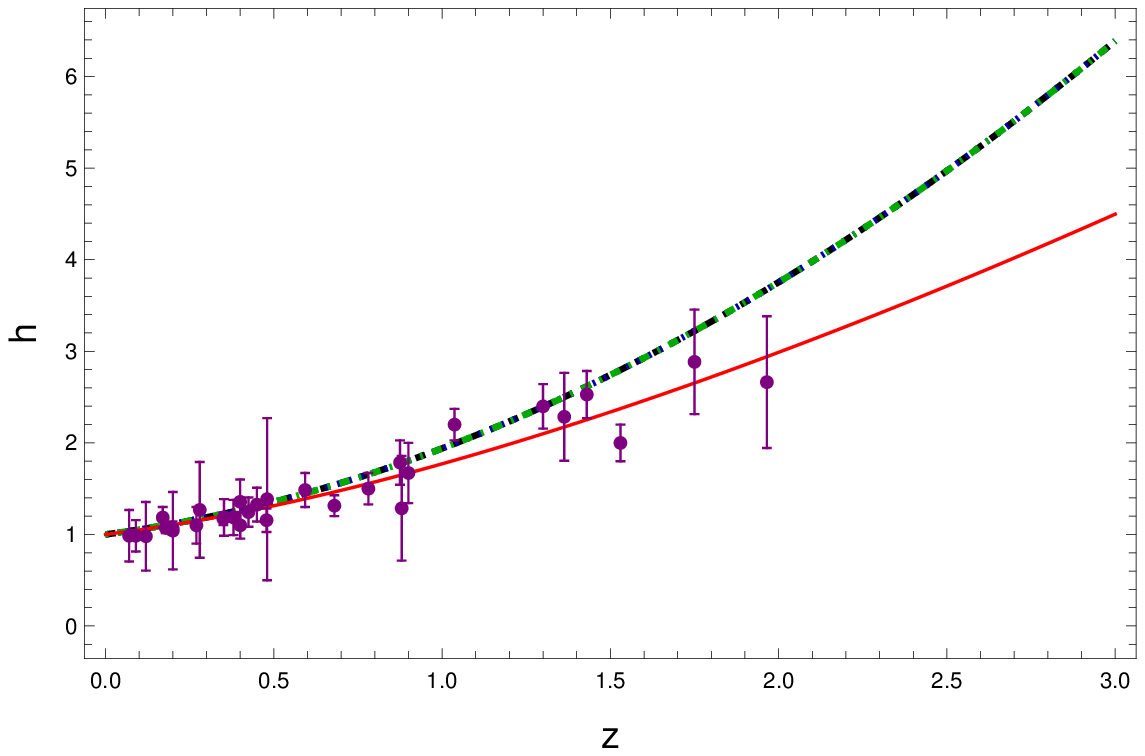}\quad\includegraphics[scale=0.7]{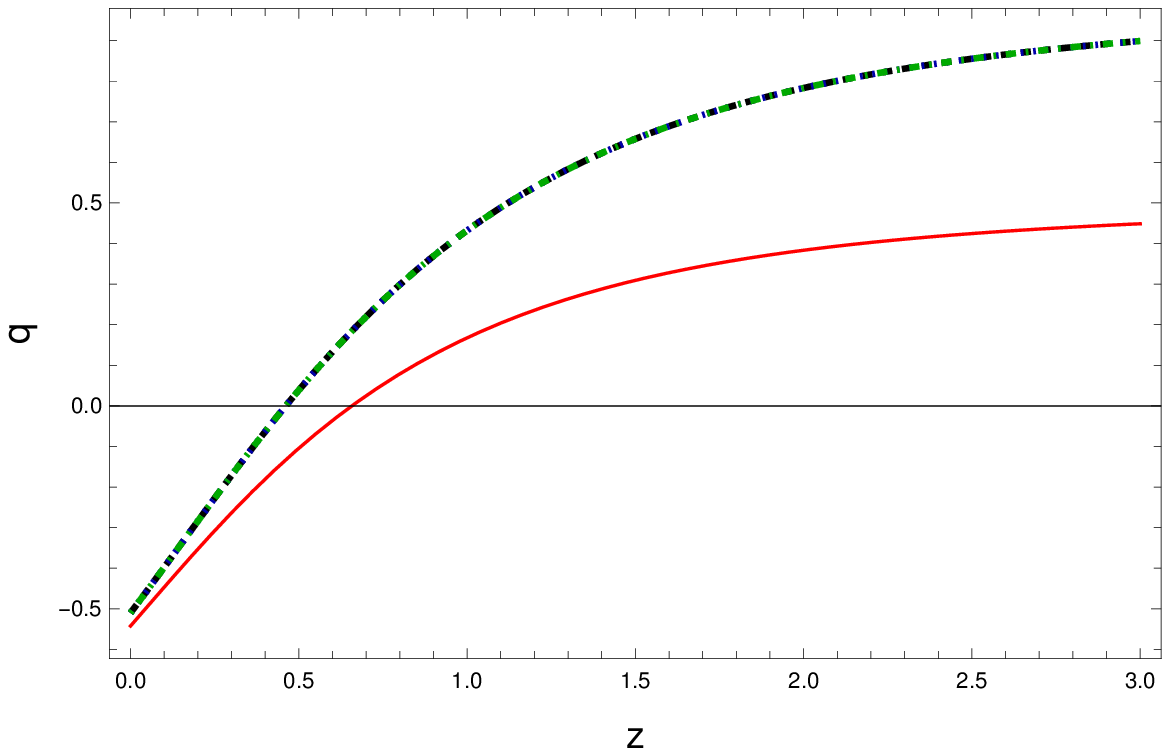}
	\caption{\label{figcase21}The Hubble function $h(z)$ and the deceleration parameter $q(z)$ for the conformally invariant quadratic Weyl cosmological model (\ref{func2}), for lower $2\sigma$ limit (dashed), best fit (dot-dashed) and upper $2\sigma$ limit (dotted) of the parameters $\alpha$ and $\bar\beta$ as given in Table~\ref{tab1a}. The solid line corresponds to the $\Lambda$CDM model, and the error bars are the observational data with their errors.}
\end{figure*}
\begin{figure*}
	\includegraphics[scale=0.7]{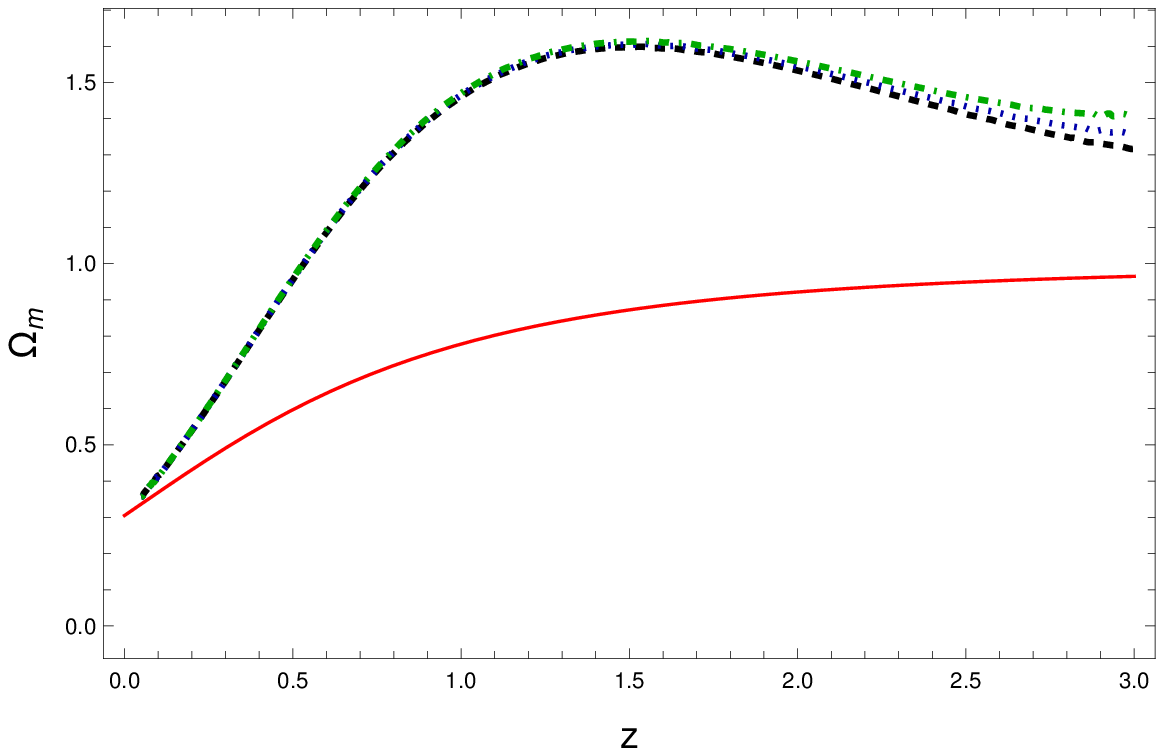}~~\quad\includegraphics[scale=0.9]{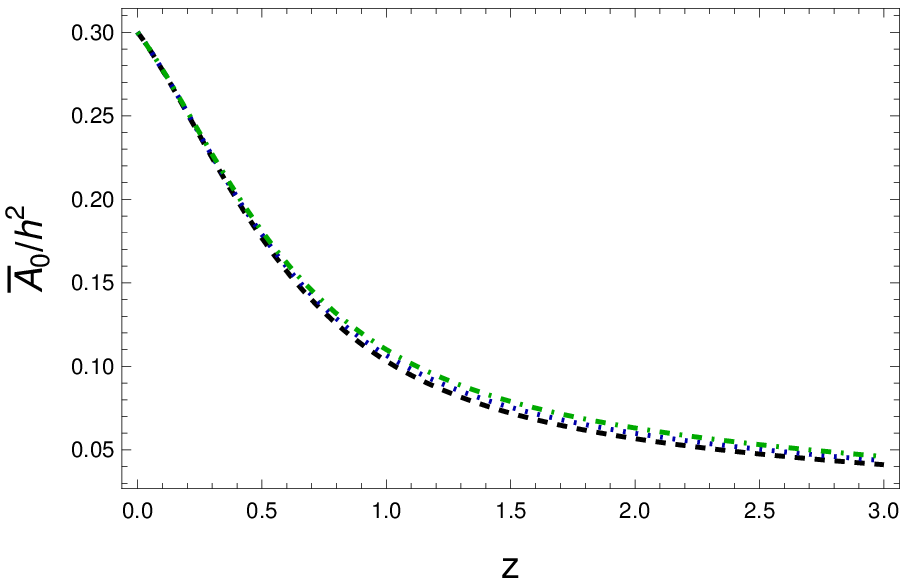}
	\caption{\label{figcase22}The redshift evolution of the dust density abundance  $\Omega_m(z)$  and of the Weyl vector $A_0(z)$ for the conformally invariant quadratic Weyl cosmological  model (\ref{func2}), for lower $2\sigma$ limit (dashed), best fit (dot-dashed) and upper $2\sigma$ limit (dotted) of the parameters $\alpha$ and $\bar\beta$ in Table ~\ref{tab1a}. The solid line corresponds to the $\Lambda$CDM theory.}
\end{figure*}

In Figs.~\ref{figcase21} and \ref{figcase22} we have plotted the evolution of the Hubble function $h$ and of the deceleration parameter $q$, together with the dust abundance $\Omega_m=\bar\rho/h^2$, and the re-scaled temporal component of the Weyl vector $\bar{A}_0/h^2$, as functions of the redshift coordinate $z$.

We have assumed the lower  $2\sigma$ limit (dashed), best fit (dot-dashed) and upper $2\sigma$ limit (dotted) of the parameters $\alpha$ and $\bar\beta$ as given in Table ~\ref{tab1a}. Also, the initial conditions we have chosen are $h^\prime(0)=0.45$, $h^{\prime\prime}(0)=0.75$, $\bar{A}_0(0)=0.3$ and $\bar{A}_0^\prime(0)=0.2$, respectively. It should be noted that the value of the parameter $\bar\gamma$ is chosen in such a way that the current value of the dust density abundance becomes equal to its $\Lambda CDM$ value $\Omega_{m0}=0.305$.

As one can see from Figs.~\ref{figcase21}, the conformally invariant geometric Weyl model can give an acceptable description of the observational data, and it reproduces the predictions of the $\Lambda$CDM model at low redshifts. However, very important differences do appear in the matter behavior, both on a quantitative and qualitative level, which may indicate the presence of very serious discrepancies between the present model and $\Lambda$CDM. The temporal component of the Weyl vector is a monotonically decreasing function of the redshift, and its evolution does not depend significantly on the variation in the numerical values of the model parameters.

\subsection{Cosmology of the conformally invariant quadratic Weyl geometric model}

We consider now the general conformally invariant Weyl geometric cosmological model, with for which the functions $f_1$, $f_2$ and $f_3$ are given by
\begin{align}\label{func3}
f_1=\frac{1}{12\xi^2}R^2, f_2=-\frac{1}{8\xi^2}R, G=1-\frac{\xi^2}{\gamma ^2}L_m.
\end{align}

In the following, we will assume that the matter Lagrangian has the form $L_m=p$. For a dust dominated universe, one can obtain the Raychaudhuri and Weyl equations as
\begin{align}\label{eq1}
4\dddot{H}&+36H\ddot{H}+8\dot{H}(3H^2+\dot{H})+4\dot{H}^2+48H^2\dot{H}\nonumber\\&-4A_0^2\dot{H}+10HA_0\dot{A}_0-6H^2A_0^2\nonumber\\&+2A_0\ddot{A}_0+2\dot{A}_0^2-9\xi^2A_0^4=0,
\end{align}
and
\begin{align}\label{eq2}
\ddot{A}_0+3(H\dot{A}_0+A_0\dot{H})+\frac{3}{2\xi^2}(2\dot{H}+H^2)A_0+9A_0^3=0,
\end{align}
respectively. Similarly to the previous case, the Friedmann equation is algebraic with respect to the matter energy density, and it could be used to obtain $\rho$ in terms of the Hubble function, and $A_0$. Now, let us define the set of dimensionless variables,
\begin{align}
&t=H_0\tau,\quad H=H_0 h,\quad A_0=H_0\bar{A}_0\nonumber\\ &\bar\rho_m=\frac{\rho_m}{6\kappa^2H_0^2},\quad \bar\gamma=\frac{\gamma}{3H_0\kappa\xi^2},
\end{align}
where $H_0$ is the current value of the Hubble parameter and $\kappa^2=1/16\pi G$. As we have already pointed out in the previous Section, the coupling constant $\xi$ is dimensionless. Transforming to the redshift coordinate $z$, one can solve numerically the system of equations~\eqref{eq1} and \eqref{eq2}, respectively,  to obtain the evolution of $h$ and $\bar{A}_0$.

In order to find the best fit value of the parameter $\epsilon$, we use again the Likelihood analysis, using the observational data on the Hubble parameter in the redshift range $z\in(0,2)$ \cite{hubble}. The best fit values of the model parameter $\xi$ and the Hubble parameter $H_0$ are summarized in Table~\ref{tab2b}.
\begin{table}
	\begin{tabular}{|c||c||c||c|}
		\hline
		Parameter&Best fit&$1\sigma$ C.I.&$2\sigma$ C.I.\\
		\hline\hline
		$\xi$&-1.77&$-1.77\pm0.38$&$-1.77\pm0.76$\\
		\hline
		$H_0$&$70.91$&$70.91\pm1.48$&$70.91\pm2.91$\\
		\hline
	\end{tabular}
\caption{The best fit values of the parameter $\xi$ and of the present day value of the Hubble function $H_0$ for the general conformally invariant quadratic Weyl model (\ref{func3}).}\label{tab2b}
\end{table}

 \begin{figure*}
	\includegraphics[scale=0.7]{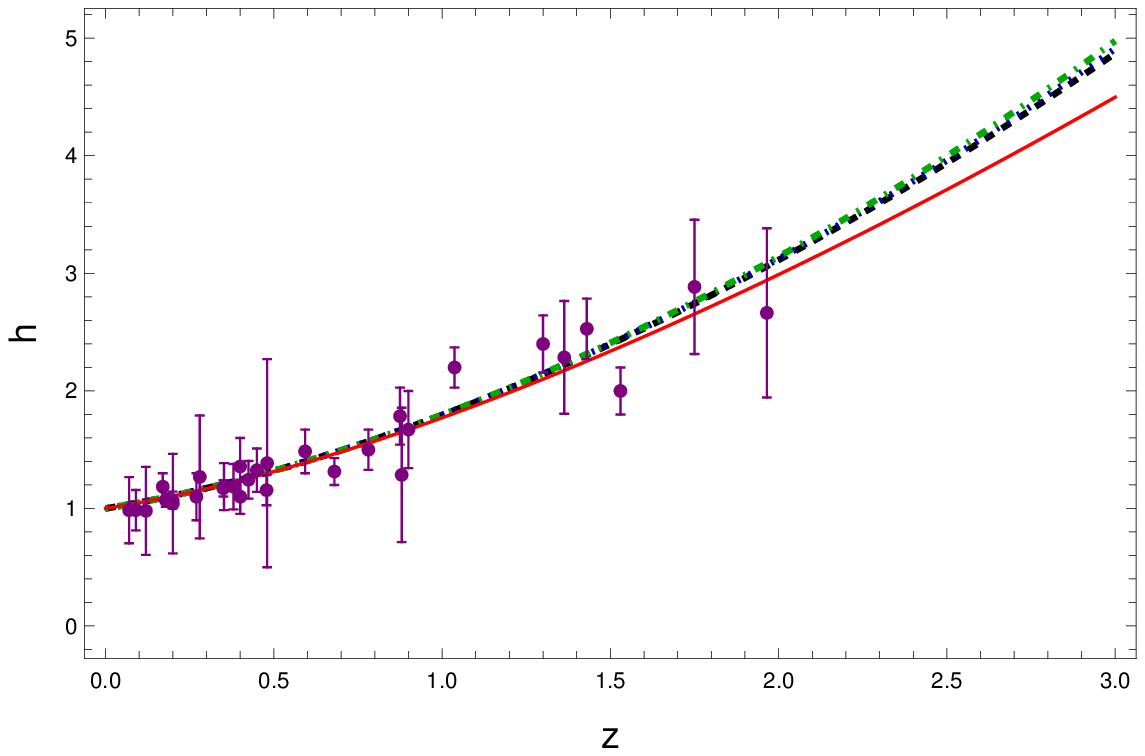}\quad\includegraphics[scale=0.7]{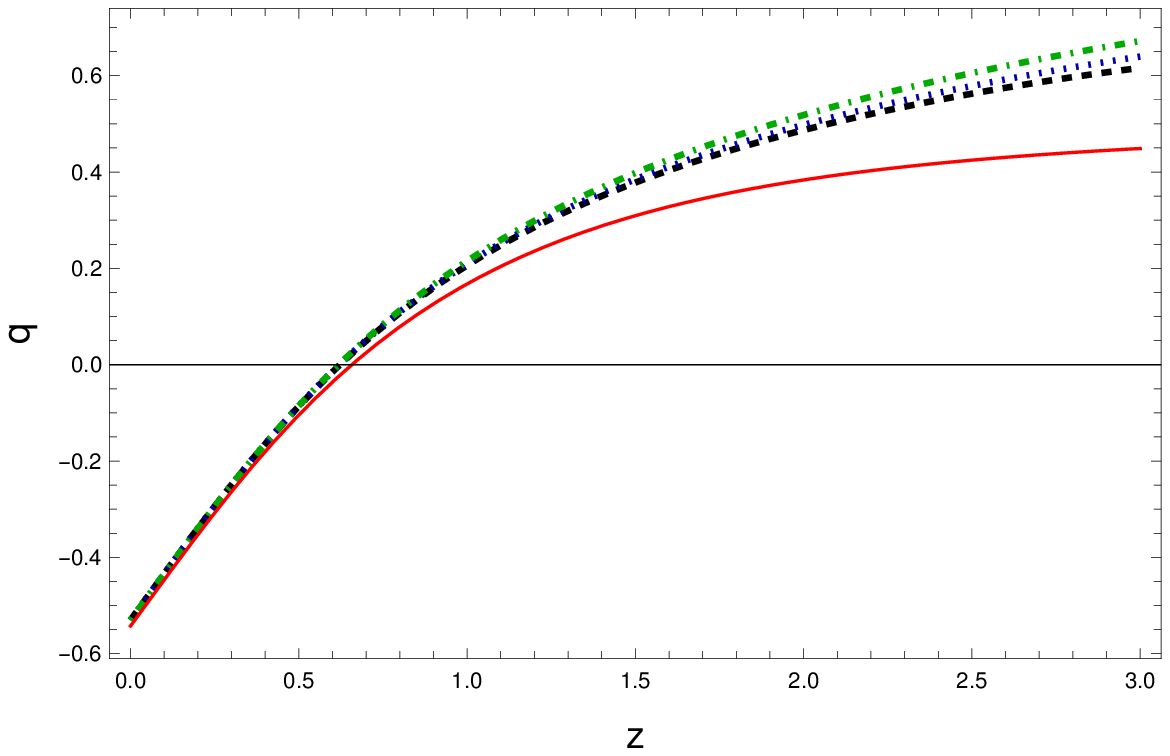}
	\caption{\label{fig3}The Hubble function $h(z)$ and the deceleration parameter $q(z)$  for the general quadratic conformally invariant Weyl geometric model (\ref{func3}), for lower $2\sigma$ limit (dashed), best fit (dot-dashed) and upper $2\sigma$ limit (dotted) of the parameter $\xi$  in the Table~\eqref{tab2b}. The solid line corresponds to the $\Lambda$CDM model, and the error bars are the observational data with their errors.}
\end{figure*}

\begin{figure*}
	\includegraphics[scale=0.7]{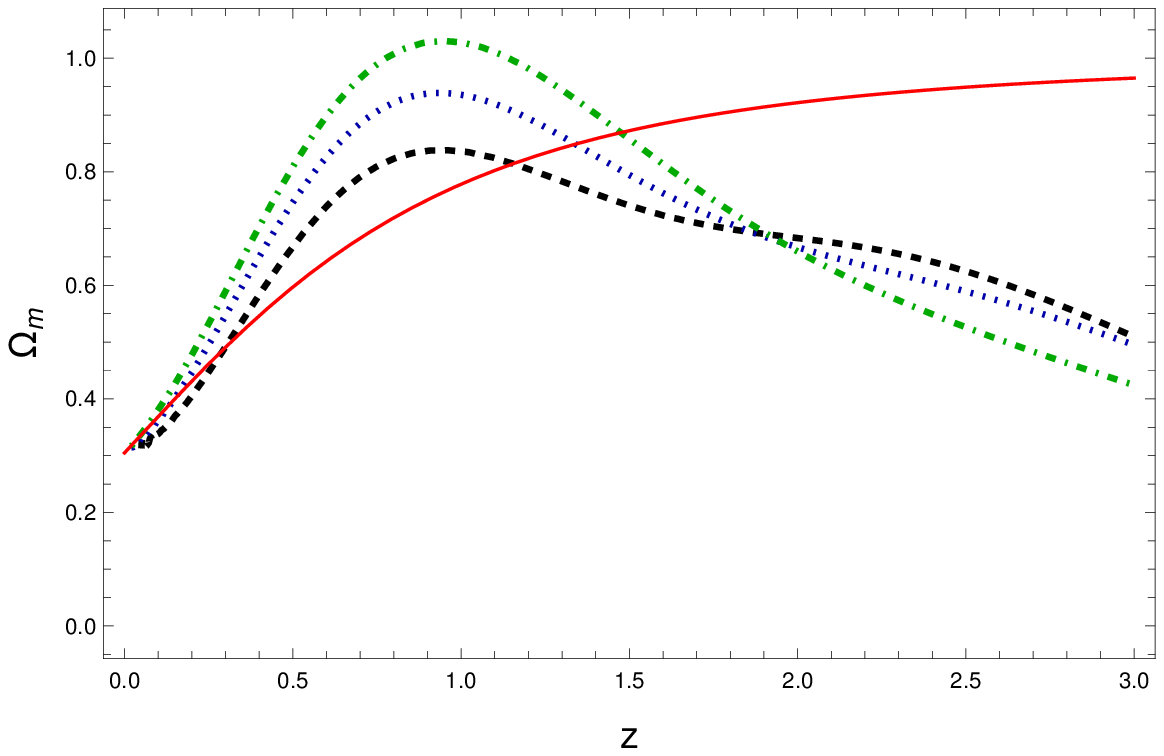}~~~\quad\includegraphics[scale=0.93]{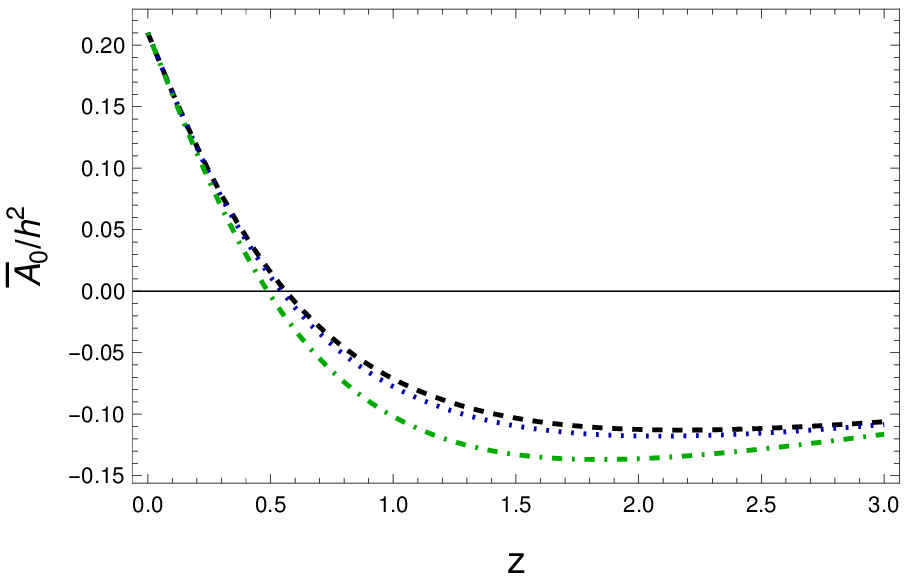}
	\caption{\label{fig4}The dust density abundance $\Omega_m(z)$ (left panel) and the re-scaled temporal component of the Weyl vector $A_0(z)$ (right panel) for the general conformally invariant Weyl geometric case (\ref{func3}), for lower $2\sigma$ limit (dashed curve), best fit (dot-dashed curve) and upper $2\sigma$ limit (dotted curve) of the parameter $\xi$ as given in Table~\ref{tab2b}. The solid line corresponds to the $\Lambda$CDM model.}
\end{figure*}

In Figs.~\ref{fig3} and \ref{fig4} we have plotted the evolution of the Hubble function $h(z)$ and of the deceleration parameter $q(z)$, together with the dust abundance $\Omega_m=\bar\rho/h^2$, and the re-scaled temporal component of the Weyl vector $\bar{A}_0/h^2$, as a function of the redshift coordinate.

In the Figs.~\ref{fig3} and \ref{fig4} we have presented the lower $2\sigma$ limit (dashed curve), the best fit (dot-dashed curve), and the upper $2\sigma$ limit (dotted curve) of the parameter $\xi$, as given in Table~\ref{tab2b}. The initial conditions we have chosen are $h^\prime(0)=0.47$, $h^{\prime\prime}(0)=0.71$, $\bar{A}_0(0)=0.21$ and $\bar{A}_0^\prime(0)=-0.3$, respectively.

It should be noted that the value of the parameter $\bar\gamma$ is chosen in such a way that the current value of the dust density abundance becomes equal to its $\Lambda CDM$ value $\Omega_{m0}=0.305$.

Similarly to the previous cases, the model gives a good description of the behavior of the Hubble function and of the observational data with respect to both observations, and the $\Lambda$CDM model. Significant differences do appear at higher redshifts of the order of $z\approx 3$. The evolution of the matter density parameter $\Omega_m$ is significantly different from its behavior in the $\Lambda$CDM model, on both quantitative and qualitative levels. The matter density reaches a maximum value at a redshift of $z\approx 0.9$, and during the different phases of the evolution the matter density can be both a time increasing, or a time decreasing function.

Thus, further precise observations of the matter density behavior at higher redshifts may provide a powerful test of the validity of the confomally invariant Weyl type cosmological models with curvature-matter coupling.

\section{Discussions and final remarks}\label{sect4}

In the present paper we have investigated the Palatini version of a generalized $f\left(R,L_m\right)$ type gravity model, originating in the conformally invariant gravity theory proposed by Weyl more than one hundred years ago, and based on his generalization of the Riemann geometry. The present theory, constructed ab initio from Weyl geometry, contains a supplementary curvature matter term, as compared to the original Weyl theory, and most of its extensions. This term can be added in a conformally invariant way to the gravitational action, so that the conformal invariance of the theory is not broken.

The requirement of the conformal invariance of the physical laws is a fundamental concept in theoretical physics, as initially suggested by Weyl \cite{Weyl1,Weyl2,Weyl3,Weyl4}. A highly attractive idea,  conformal invariance is analogous to the gauge principle in the physics of elementary
particle, where it played a fundamental role in the advancement of modern physics. There are strong similarities between the global transformations of units, and the global gauge transformations. The laws governing the realm of the elementary particle physics are conformally invariant, which is not the case for Einstein’s gravity.
This aspect represent another important difference between microphysics, and the gravitational interaction. A bridge between elementary particle physics and gravitation can be constructed via the Weyl geometry, which naturally contains the principle of conformal invariance. The simplest fully conformally invariant theory of gravity contains in its action a quadratic term in the Weyl scalar, as well as the contribution coming from the strength of the Weyl field. Once a conformally invariant matter Lagrangian term is added, the resulting theory is fully conformally invariant. However, generally the matter Lagrangian is not conformally invariant, and to assure the conformal invariance of the theory in presence of matter a coupling of the matter Lagrangian to curvature is also necessary. The metric version of such a theory was constructed and investigated in \cite{HaSh}, where its cosmological implications have also been considered. In the present paper we have extended the Weyl geometric $f\left(R,L_m\right)$ theory by assuming arbitrary couplings between matter and Weyl geometric terms, couplings that may not guarantee the conformal invariance of the considered models. However, even in the presence of a broken conformal invariance, theoretical gravity models based on Weyl geometry may prove relevant for the interpretation of the present day observations.

To construct a $f\left(R,L_m\right)$ type Weyl geometric model we begin with the simplest possible conformal invariant action, containing the square of the Weyl scalar, the strength of the Weyl vector, and a matter-curvature coupling term. Due to the straightforward relation existing between the geometric terms in Weyl and Riemann geometries, Weyl models can be equivalently reformulated in Riemann geometry. In the case of the quadratic Weyl action the equivalent gravitational action in the Riemann geometry contains a term proportional to $R^2$, coupled with matter, a term containing the Ricci scalar, multiplied by $\omega ^2$, and also coupled with matter, plus some other terms originating in Weyl geometry. In our analysis we have generalized this model, by considering arbitrary functional couplings instead of the $R^2$ and $R$ terms in the action.

To obtain the field equations we have considered the independent variation of the action with respect to the metric and the independent connection. As a first result of these variations it follows that the independent connection $\bar{\Gamma}$ is compatible with a conformal metric $h_{\mu \nu}$, given by $h_{\mu \nu}=F\left(R,\omega ^2,L_m\right)g_{\mu \nu}$, with the conformal factor $F$ a function of the Ricci scalar, the matter Lagrangian, and the square of the Weyl vector. The metric $h_{\mu \nu}$ is assumed now to be the physical metric, and the associated, Levi-Civita type connection can be represented in Riemann geometry as the sum of the Levi-Civita connection of the metric $g$ plus corrections terms coming from the conformal factor $F$. Once the form of the connection is known, the geometric quantities and the Einstein field equations can be easily obtained, thus leading to the Palatini formulation of the generalized Weyl geometric gravity in the presence of curvature-matter coupling.

As a possible test of the Palatini formulation of the Weyl geometric type $f\left(R,L_m\right)$ theories we have investigated their cosmological implications. The cosmological evolution equation (generalized Friedmann equations) can be obtained in their general We have considered three specific models, based on particular choices of the functions $f_1$, $f_2$ and $G$ that determine the mathematical structure of the theory. The first (toy) model corresponds to an action of the form
\be\label{act1}
S=\int{\left[\frac{1}{2}\alpha \gamma +\beta \gamma R\omega ^2+\frac{9}{4}\gamma \omega ^4-\frac{1}{4}\tilde{F}_{\mu\nu}^2\right]\sqrt{-g}d^4x},
\ee
in which the matter Lagrangian has been approximated by a constant. The corresponding cosmological model admits a de Sitter solution, corresponding to a constant temporal component of the Weyl vector. A second class of solutions obtained from the action (\ref{act1}) correspond to models with varying Hubble function, having fixed values of $A_0$ and $\gamma$. Despite their simplicity, these models can give an acceptable description of the observational data for the Hubble function, and can reproduce rather well the behavior of the Hubble function up to redshifts of $z\approx 1$.

The second (toy) cosmological model we have considered is derived from the action
\bea
S&=&\int\Bigg[\frac{1}{2}\alpha\left(1-\gamma L_m\right)R^2+\beta \omega ^2+\frac{9}{4}\left(1-\gamma L_m\right)\omega ^4\nonumber\\
&&-\frac{1}{4}\tilde{F}_{\mu\nu}^2\Bigg]\sqrt{-g}d^4x,
\eea
with $\alpha$, $\beta$ and $\gamma $ constants. The Palatini version of this model leads to third order differential equations for the evolution of the Hubble function. However, the model can describe well the low redshift observations, but the differences between the predictions of this model and those of the $\Lambda$CDM model significantly increase with the redshift. The behavior of the matter density parameter is very different as compared to the $\Lambda$CDM model, suggesting the existence of periods in which the matter density increases, and decreases, respectively.

Finally, the third cosmological model we have considered is based on the Palatini variation of the Weyl geometric action
\bea
S&=&\int \Bigg[\frac{1}{24\xi^2}\left(1-\frac{\xi^2}{\gamma^2}L_m\right)R^2-\frac{1}{8\xi^2}\left(1-\frac{\xi^2}{\gamma ^2}L_m\right)\omega ^2R\nonumber\\
&&+\frac{9}{4}\left(1-\frac{\xi^2}{\gamma^2}L_m\right)\omega ^4-\frac{1}{4}\tilde{F}^2_{\mu \nu}\Bigg]\sqrt{-g}d^4x,
\eea
with $\xi$ and $\gamma$ constants. This model also gives third order ordinary differential evolution equations for the Hubble function. The fitting with the observation al data allows the determination of the model parameters. There is a good concordance with observational data for the Hubble function up to a redshift of around $z\approx 2$. The differences between models increase with increasing redshift. For low redshifts ($z<0.5$) the matter density parameter of the model can reproduce the $\Lambda$CDM behavior.

The temporal component $A_0$ of the Weyl vector has a similar behavior in all three considered models, being a monotonically decreasing function of redshift (a monotonically increasing function of time), indicating that it had much higher values in the early Universe. For the third considered model the Weyl vector takes negative values for redshifts larger than 0.5.

To conclude, the results of our investigations performed in the present work suggest that the Palatini formulation of the Weyl geometric $f\left(R,L_m\right)$ type theories of gravity may play a relevant role in the description of gravitational processes at both high matter densities, corresponding to the very early Universe, and at the low densities, specific to the present day Universe. Other astrophysical and cosmological implications of the Palatini formulation of Weyl geometric type theories will be considered in a future study.

\section*{Acknowledgments}

The work of T.H. is supported by a grant of the Romanian Ministry of
Education and Research, CNCS-UEFISCDI, project number
PN-III-P4-ID-PCE-2020-2255 (PNCDI III).

\end{document}